\documentclass[a4paper,12pt]{article}
\usepackage{epsfig}
\newcommand{\bmat}{\left(\begin{array}}
\newcommand{\emat}{\end{array}\right)}
\def\NPB#1#2#3{Nucl. Phys. B{#1} (19#2) #3}

\def\PRD#1#2#3{Phys. Rev. D{#1} (19#2) #3}

\def\lsim{\raise0.3ex\hbox{$\;<$\kern-0.75em\raise-1.1ex\hbox{$\sim\;$}}}
\def\gsim{\raise0.3ex\hbox{$\;>$\kern-0.75em\raise-1.1ex\hbox{$\sim\;$}}}

\def\yzero{\smash{\hbox{$y\kern-4pt\raise1pt\hbox{${}^\circ$}$}}}

\def\s2{\frac{1}{\sqrt2}}

\def\beq{\begin{equation}}
\def\eeq{\end{equation}}
\def\beqa{\begin{eqnarray}}
\def\eeqa{\end{eqnarray}}

\def\IF{\relax{\rm I\kern-.18em F}}
\def\II{\relax{\rm I\kern-.18em I}}
\def\IP{\relax{\rm I\kern-.18em P}}
\def\IC{\relax\hbox{\kern.25em$\inbar\kern-.3em{\rm C}$}}
\def\IR{\relax{\rm I\kern-.18em R}}

\def\Dsl{\,\raise.15ex\hbox{/}\mkern-13.5mu D} 
\def\IZ{Z\kern-.4em  Z}
\def\bmat{\left(\begin{array}}
\def\emat{\end{array}\right)}

\def    \part          {\partial}
\def    \be            {\begin{equation}}
\def    \ee            {\end{equation}}
\def    \bea           {\begin{eqnarray}}
\def    \eea           {\end{eqnarray}}
\def    \nn            {\nonumber}

\def    \ccii          {C_i^{5_i}}

\def    \ccia          {C_1^{5_i}}  
\def    \ccib          {C_2^{5_i}}  
\def    \ccic          {C_3^{5_i}}  
\def    \ccij          {C_{j}^{5_{i}}}

\def    \ccjk          {C_{j}^{5_{k}}}

\def    \cni           {C_i^{9}}  
\def    \cna           {C_1^{9}}
\def    \cnb           {C_2^{9}} 
\def    \cnc           {C_3^{9}}
\def    \cnci          {C^{9 5_i}}
\def    \cncj          {C^{9 5_{j}}}   
\def    \cnck          {C^{9 5_{k}}}

\def    \ccjck         {C^{5_{j} 5_{k}}}
\def    \ccick         {C^{5_{i} 5_{k}}}
\def    \ccicj         {C^{5_{i} 5_{j}}}
\def    \ccacb         {C^{5_1 5_2}}
\def    \cccca         {C^{5_3 5_1}}
\def    \ccbcc         {C^{5_2 5_3}}
\begin{document}
%
\pagestyle{empty}
\rightline{FTUAM 01/03}
\rightline{IFT-UAM/CSIC-01-04}
\rightline{HIP-2000-69/TH}
\rightline{SUSX-TH/01-007}
\rightline{February 2001}

\renewcommand{\thefootnote}{\fnsymbol{footnote}}
\setcounter{footnote}{0}

\vspace{0.0cm}
\begin{center}
\large{\bf Determination of the String Scale in D-Brane Scenarios and 
Dark Matter Implications\\[5mm]}
\mbox{
\sc{
\small
{D.G. Cerde\~no$^{1,2}$,
E. Gabrielli$^{3}$,
S. Khalil$^{4,5}$,
C. Mu\~noz$^{1,2}$,
E. Torrente-Lujan$^{1}$
}
}
}
\begin{center}
{\small
{\it $^1$ Departamento de F\'{\i}sica
Te\'orica C-XI, Universidad Aut\'onoma de Madrid,\\[-0.1cm]
Cantoblanco, 28049 Madrid, Spain. \\
\vspace*{2mm}
\it $^2$ Instituto de F\'{\i}sica Te\'orica  C-XVI,
Universidad Aut\'onoma de Madrid,\\[-0.1cm]
Cantoblanco, 28049 Madrid, Spain.\\
\vspace*{2mm}
\it $^3$ Institute of Physics, University of Helsinki,\\[-0.1cm]
P.O. Box 9, Siltavuorenpenger 20 C, FIN-00014 Helsinki, Finland.\\
\vspace*{2mm}
\it $^4$ Centre for Theoretical Physics, University of Sussex, 
Brighton BN1 9QJ, U.K.\\
\vspace*{2mm}
\it $^5$ Ain Shams University, Faculty of Science, Cairo
11566, Egypt.} 
}
\end{center}

{\bf Abstract} 
\\[7mm]
\end{center}
\begin{center}
\begin{minipage}[h]{14.0cm}
We analyze different phenomenological aspects of D-brane constructions.
First, we obtain that 
scenarios with the gauge group and particle content of the
supersymmetric standard model lead naturally to intermediate values for the
string
scale, in order to reproduce the value of gauge couplings
deduced from experiments.
Second, the soft terms, which turn out to be generically 
non universal, and Yukawa couplings of
these scenarios are studied in detail.
Finally, using these soft terms and the string scale as the initial
scale for their running, 
we compute the neutralino-nucleon cross section. In particular
we find regions in the parameter space of D-brane scenarios with
cross sections in the range 
of $10^{-6}$--$10^{-5}$ pb, i.e. where current dark matter
experiments are sensitive. For instance, this can be obtained for 
$\tan\beta > 5$.




\end{minipage}
\end{center}
\vspace{0.5cm}
\begin{center}
\begin{minipage}[h]{14.0cm}

\end{minipage}
\end{center}
\newpage
\setcounter{page}{1}
\pagestyle{plain}
\renewcommand{\thefootnote}{\arabic{footnote}}
\setcounter{footnote}{0}
%
%
\section{Introduction}

Although the standard model provides
a correct description of the observable world, there exist, however,
strong indications that it is just an effective theory at low energy
of some fundamental one. The only candidates for such a theory are,
nowadays, the string theories, which have the potential to unify
the strong and electroweak interactions with gravitation in a
consistent
way.

In the late eighties, working in the context of the perturbative 
heterotic string, 
a number of interesting four-dimensional vacua
with particle content not far from that of the supersymmetric 
standard model were found \cite{viejos}. Supersymmetry breaking was
most of the times assumed to take place non-perturbatively 
by gaugino condensation
in a hidden sector of the theory. Until recently, 
it was thought that this was the only way in order to construct
realistic string models. However, in the late nineties, we have
discovered that explicit models with realistic properties 
can also be constructed using D-brane configurations from 
type I string vacua 
\cite{kaku2}-\cite{kingg}. 
Besides, it has been realized that
the string scale, $M_I$, may be anywhere between the weak scale, $M_W$,
and the Planck scale, 
$M_{Planck}$ \cite{Lykken,Dimopoulos,typeITeV,kaku,stronghete,typeIinter}. 
This is to be compared
to the perturbative heterotic string where the relation
$M_I=\sqrt{\frac{\alpha}{8}} M_{Planck}$, with $\alpha$ the gauge coupling,
fixes the value of the string scale.

The freedom to play with the value of $M_I$ is particularly interesting
since there are several arguments in favour of supersymmetric
scenarios with scales $M_I\approx 10^{10-14}$ GeV.
First, these scales were suggested in \cite{stronghete}
to explain many experimental observations as
neutrino masses or the scale for axion physics. 
Second, with the string scale of order 
$10^{10-12}$ GeV
one is able
to attack the hierarchy problem of unified theories \cite{typeIinter}.
The mechanism is the
following.
In supergravity models supersymmetry can be spontaneously broken in a 
hidden sector of the theory and the gravitino mass,
which sets the overall scale of the soft terms, is given by:
\bea
m_{3/2}\approx \frac{F}{M_{Planck}}\ ,
\label{gravitino}
\eea
where $F$ is the auxiliary field whose vacuum expectation value
breaks supersymmetry. 
Since in supergravity one would expect $F\approx M_{Planck}^2$, one  
obtains
$m_{3/2}\approx M_{Planck}$ and therefore 
the hierarchy problem solved in principle by supersymmetry
would be re-introduced,
unless non-perturbative effects such as gaugino condensation
produce $F\approx M_W M_{Planck}$. However, if the scale
of the fundamental theory is $M_I\approx 10^{10-12}$ GeV instead of
$M_{Planck}$,
then $F\approx M_I^2$
and one gets $m_{3/2}\approx M_W$ in a natural way, without invoking any
hierarchically suppressed non-perturbative effect.
Third, for intermediate scale scenarios 
charge and color breaking constraints become less important.
Let us recall that
charge and color breaking minima in supersymmetric theories might make 
the standard vacuum unstable.
Imposing that the standard vacuum should be the global minimum
the corresponding
constraints
turn out to be very strong and, in fact, working with the usual
unification scale $M_{GUT}\approx 10^{16}$ GeV, 
there are extensive regions in the
parameter
space of soft supersymmetry-breaking terms that become
forbidden \cite{mua}. For example, for the dilaton-dominated scenario
of superstrings the whole parameter space turns out to be excluded 
\cite{mua3} on these grounds.
The stability of the corresponding constraints with respect to
variations
of the initial scale for the running of the soft breaking terms
was studied in \cite{mua2}, finding that the larger the scale is,
the stronger the bounds become. In particular, by taking $M_{Planck}$
rather than $M_{GUT}$ for the initial scale stronger constraints were
obtained. Obviously the smaller the scale is, the weaker the 
bounds become. In \cite{abel} intermediate scales rather than 
$M_{GUT}$ were considered for the dilaton-dominated scenario with
the interesting result that it is allowed in a large region of
parameter
space.
Finally, 
there are other arguments in favour 
of scenarios with intermediate string scales $M_I\approx 10^{10-14}$
GeV.
For example these scales
might also
explain the observed ultra-high energy ($\approx 10^{20}$ eV) cosmic rays
as products of long-lived massive string mode decays. Besides,
several models of chaotic inflation favour also these scales \cite{caos}.

In the present article we are going to analyze in detail
whether or not those intermediate string scales are also 
necessary in order to reproduce the
low-energy data, i.e. the values of the gauge
couplings deduced from CERN $e^{+}e^{-}$ collider LEP experiments.
In this sense, we will see that D-branes scenarios indeed
lead naturally to intermediate values for the
string scale $M_I$.

On the other hand, it has been noted that 
the neutralino-nucleon cross section 
is quite sensitive to the value of the initial scale for the running
of the soft breaking terms \cite{Nosotros}. The smaller the scale is, the
larger the cross section becomes. In particular, by taking
$10^{10-12}$ GeV rather than $10^{16}$ GeV
for the initial scale, 
the cross section increases substantially 
$\sigma\approx 10^{-6}$--$10^{-5}$ pb.
This result is extremely interesting since
the lightest neutralino is usually the lightest supersymmetric particle 
(LSP), and therefore
a natural candidate
for dark matter in supersymmetric theories \cite{review},
and current dark matter detectors,
DAMA \cite{experimento1} and CDMS \cite{experimento2},
are sensitive to a neutralino-nucleon cross section in the above range.

The initial scale for the running of the soft terms in D-brane scenarios
is $M_I$. As mentioned above, several theoretical and phenomenological
arguments suggest that intermediate values for this scale
are welcome. Thus 
it is natural to wonder 
how much the standard neutralino-nucleon
cross section analysis will get modified 
in D-brane scenarios.
This is another aim of this article.

The content of the article is as follows. In Section 2 
we will try to determine the string scale in D-brane scenarios
imposing
the experimental constraints on the values of the gauge coupling
constants. Although we will concentrate mainly in scenarios
where the $SU(3)_c$, $SU(2)_L$ and $U(1)_Y$ groups
of the standard model come from different sets of D$p$-branes,
we will also review the scenario where they come from
the same set of D$p$-branes.
The fact that the $U(1)_Y$ group arises
as a linear combination of different $U(1)$'s, due to their
D-brane origin, is crucial in the analysis.

In Section 3 we will use the results of Section 2, in particular
the matter distribution
due to the D-brane origin of the $U(1)$
gauge groups, in order to derive
the soft supersymmetry breaking terms of the D-brane scenarios which
may give rise to the supersymmetric standard model. 
Generically they are non-universal.
This analysis is carried out under the
assumption of dilaton/moduli supersymmetry breaking 
\cite{BIM22}-\cite{BIM}.
We emphasize that this assumption of 
dilaton/moduli dominance is more compelling in the D-brane scenarios 
where only closed string fields like $S$ and $T_i$ can move into 
the bulk and transmit supersymmetry
breaking from one D-brane sector to some other. Finally,
we will also discuss the structure of Yukawa coupling matrices.

In Section 4, using the soft terms of the D-brane scenarios previously
studied,
we compute the neutralino-nucleon cross section.
We will see how 
the compatibility of regions in the parameter space of these scenarios
with the sensitivity of current dark matter experiments
depends not only on the value of the string scale
but also on the non-universality of the soft terms.

Finally, the conclusions are left for Section 5.

\section{D-brane scenarios and the string scale}

As mentioned in the Introduction there exists the interesting
possibility
that the supersymmetric 
standard model might be built using 
D-brane configurations. 
In this case there are two possible avenues to carry it out:
i) The $SU(3)_c$, $SU(2)_L$ and $U(1)_Y$ groups
of the standard model come from different sets of D$p$-branes. 
ii) They come from the same set of D$p$-branes. 

Since the two
scenarios are interesting and qualitatively different, we will discuss
both separately. 
We will see in detail below that the first one (i), 
in order to reproduce 
the values of the gauge
couplings deduced from CERN $e^{+}e^{-}$ collider LEP experiments,
leads naturally to intermediate values for the
string scale $M_I$. One realizes that this is an interesting result
since there are several arguments in favour of intermediate scales,
as discussed in the Introduction.
This approach was used first in \cite{Antoniadis} for the case of
non-supersymmetric D$p$-branes with the result of a string scale
of the order of a few TeV.
In any case, it is worth remarking
the difficulty of  
obtaining three copies of quarks and leptons
if the gauge groups are attached to different
sets of D$p$-branes\footnote{
We thank L.E. Ib\'a\~nez for discussions about this point (see also
\cite{Ibanez})}.
Thus
whether or not the scenarios discussed below,
may arise from different sets of D$p$-branes 
in explicit string constructions
is an important issue which is worth attacking in the future.

Concerning the other scenario (ii), models 
with the gauge group of the standard model and
three families of particles have been explicitly built \cite{Ibanez,Bailin}.
We will review whether or not intermediate scales 
arise naturally.

\subsection{Embedding the gauge groups within different sets of D$p$-branes}

It is a plausible situation to assume that the $SU(3)_c$ and the 
$SU(2)_L$ groups of the standard model could come from different sets of
D$p$-branes \cite{ibanez2,typeITeV}. By different sets we mean D$p$-branes 
whose world-volume is not identical. In particular, notice that the 
standard model 
contains particles (the left-handed quarks $Q_u$) transforming 
both under $SU(3)_c$ and $SU(2)_L$. That means that there must be
some overlap of the world-volumes of both sets of D$p$-branes.
Thus e.g., one cannot put $SU(3)_c$ inside a set of D$3$-branes 
and $SU(2)_L$ within another set of parallel D$3$-branes on a different point
of the compact space since then there would be no massless
modes corresponding to the exchange of open strings between 
both sets of branes which could give rise to the left-handed
quarks.  

Thus we need to embed $SU(3)_c$ 
inside D-branes, say D$p_3$-branes, and $SU(2)_L$ within other
D-branes, say D$p_2$-branes,
in such a way that their corresponding world-volumes have some overlap.
Since we are working in general 
with type IIB orientifolds, $p_N$ can be 
$3$, $5_i$, $7_i$, and $9$, where the index $i=1,2,3$ denotes
what complex compact coordinate is included in the D$5$-brane
world-volume,
or is transverse to the D$7$-brane world-volume.
Not all types of D$p_{N}$-branes may be present simultaneously if
we want to preserve $N=1$ in $D=4$. For a given $D=4$, $N=1$ vacuum
we can have at most either D$9$-branes with D$5_i$-branes 
or D$3$-branes with D$7_i$-branes.

\begin{figure}[ht]
\begin{center}
\begin{tabular}{c}
\epsfig{file= 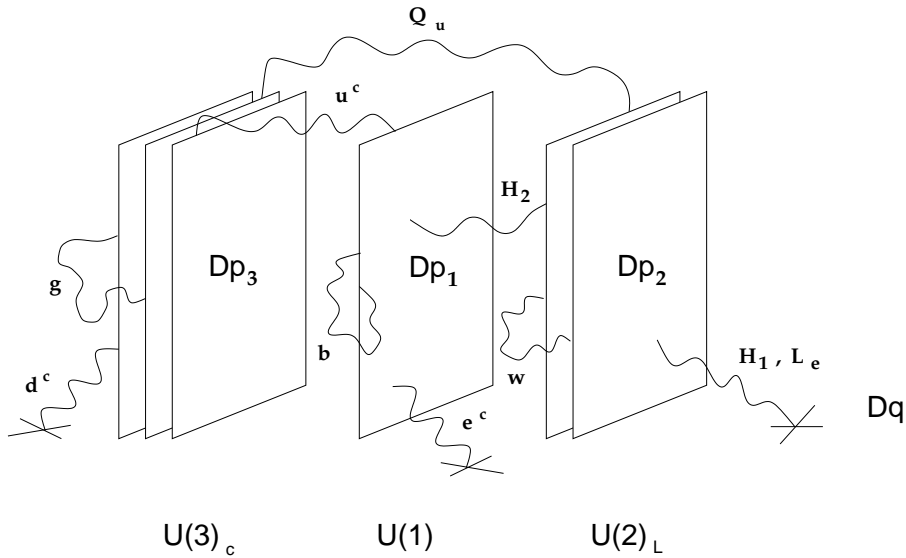, width=12cm}\\
\end{tabular}
\end{center} 
\caption{A generic D-brane scenario giving rise to the gauge bosons
and matter of the standard model. It contains three D$p_3$-branes,
two D$p_2$-branes and one D$p_1$-brane, where $p_N$ may be
either 9 and $5_i$ or 3 and $7_i$. The presence of extra D-branes,
say D$q$-branes, is also
necessary as explained in the text. For each set the D$p_N$-branes
are in fact on the top of each other.} 
\end{figure}

In type IIB orientifold models, and in general on the world-volume of
D-branes, $SU(N)$ groups come along with a $U(1)$ factor,
say $U(1)_N$, so that indeed
we are dealing with $U(N)$ groups, in which both 
$SU(N)$ and $U(1)_N$ share the same coupling constant, $\alpha_N$.
Thus $U(1)_Y$ might be a linear combination 
of two $U(1)$ gauge groups arising from $U(3)_c$ and $U(2)_L$ 
within D$p_3$- and D$p_2$-branes respectively \cite{Rigolin2}.
Although this is the simplest possibility, its analysis
is somehow subtle \cite{Antoniadis} and we prefer to carry it out at    
the end of this subsection.
Thus we will analyze first a more general case, where
an extra $U(1)$ arising from another D-brane, say D$p_1$-brane,
contributes to the combination giving rise to the correct hypercharge
of the standard model matter \cite{Antoniadis,Ibanez}. 
This is schematically shown in Fig.~1, where open strings starting and
ending on the same sets of D$p_N$-branes give rise to the gauge bosons
of the standard model. For the sake of visualization each set is
depicted at parallel locations, but in fact they are intersecting each
other as discussed above.

In \cite{Ibanez} a $Z_3$ orientifold model with 
$U(3)_c\times U(2)_L\times U(1)$ observable gauge group, and
therefore giving rise to 
$SU(3)_c\times SU(2)_L\times U(1)_3\times U(1)_2\times U(1)_1$ 
as discussed above, was explicitly built.
Nevertheless, this model is embedded in D3-branes, i.e. $p_3=p_2=p_1=3$, 
and therefore we will discuss it in detail in Subsection 2.2.

On the other hand, in \cite{Antoniadis} the existence of standard models
coming from different sets of 
{\it non-supersymmetric} D$p$-branes was assumed and several 
consequences were discussed. In particular, imposing D$p_3$=D$p_1$,
i.e. $\alpha_3=\alpha_1$,
the low-energy data are reproduced for a string scale of the order of
a few TeV. 
Here we will carry out the general analysis of {\it supersymmetric}
D$p$-branes
with the interesting result
that intermediate values ($\approx 10^{10-12}$ GeV) for the string scale
may arise in a natural way. 

\subsubsection{General scenario with $Dp_3\neq Dp_2\neq Dp_1\neq Dp_3$}

Let us denote by $Q_3$, $Q_2$ and $Q_1$ the charges of
$U(1)_3$, $U(1)_2$, and $U(1)_1$ respectively. 
Following then the analysis of  
Antoniadis, Kiritsis and Tomaras \cite{Antoniadis}
a family of quarks and leptons can have the four assignments
of quantum numbers given in Table~1, in order to obtain the 
hypercharge of the standard model 
\bea
Y=c_3\sqrt 6 Q_3+c_2\sqrt 4 Q_2+\sqrt 2 Q_1\ ,
\label{hypercharge}
\eea
where $c_3=-1/3$, 
$c_2=-1/2$ ($c_2=1/2$) for the first (second) assignment and $c_3=2/3$,
$c_2=-1/2$ ($c_2=1/2$) for the third (fourth) assignment.
Note that $U(N)$ generators are normalized as 
Tr $T^aT^b=\frac{1}{2}\delta^{ab}$, and therefore the
fundamental representation of $SU(N)$ has
$Q_N=1/\sqrt{2N}$. 
\begin{table}
\begin{center}
\begin{tabular}[htb]{|c|c|cc|cc|cc|cc|c|}
\hline
Matter Fields&$Q_3$&$Q_2$&$Q_1$&$Q_2$&$Q_1$&$Q_2$&$Q_1$&$Q_2$&$Q_1$&Y\\
\hline
\hline
$Q_u(3,2)$     &1 &-1&0 &1 &0 &1 &0 &-1&0 &1/6 \\
$u^c(\bar 3,1)$ &-1&0 &-1&0 &-1&0 &0 &0 &0 &-2/3 \\
$d^c(\bar 3,1)$ &-1&0 &0 &0 &0 &0 &1 &0 &1 &1/3 \\
$L_e(1,2)$     &0 &1 &0 &1 &-1&1 &0 &1 &-1&-1/2 \\
$e^c(1,1)$      &0 &0 &1 &0 &1 &0 &1 &0 &1 &1\\
\hline
\end{tabular}
\caption{
The four possible assignments of 
quantum numbers (multiplied by $\sqrt{2N}$) 
of a family of quarks and leptons
of the standard model under $U(1)_3\times U(1)_2\times U(1)_1$.
Note that $Q_3$ is always fixed.
The usual hypercharge $Y$ is given in the last column.
}
\end{center}
\end{table}

For example, as discussed above,
the quark doublet $Q_u$ always arises from an open string with one end
on D$p_3$-branes and the other end on D$p_2$-branes.
In the first assignment of Table~1 $Q_u$ transforms as a $\bar 2$
under $U(2)$ and therefore $Q_2=-1/\sqrt 4$.
$u^c$ ($d^c$) arises from an open string with one end
on D$p_3$-branes and the other end on D$p_1$ 
(D$q$)-branes\footnote{As we see from here the presence of extra D-branes,
say D$q$-branes, is necessary in order to reproduce the correct 
hypercharge for the matter. 
In addition, in Subsection 2.2 we will see an explicit model
where D$q$-branes are also necessary to cancel non-vanishing tadpoles.
The additional $U(1)$ factors associated to the 
D$q$-branes will be anomalous and therefore with a 
mass of the order of the string scale.}
with $Q_1=-1/\sqrt 2$ ($0$). Finally, in the case of leptons, 
$L_e$ ($e^c$) arises
from an open string with one end on D$p_2$ (D$p_1$)-branes and the other end
on D$q$-branes with $Q_1=0$ ($1/\sqrt 2$).
This is schematically shown in Fig.~1. 
The other three possible assignments
can also be analyzed similarly. Let us remark that other scenarios
with $u^c$, $d^c$ ($e^c$) arising from open strings with
both ends on D$p_3$ (D$p_2$)-branes 
are possible, since these particles 
can be obtained as the antisymmetric product
of two triplets of $SU(3)$ (doublets of $SU(2)$). However, these scenarios
do not give rise to a modification
of the analysis of the string scale \cite{Antoniadis}, and therefore
we will not consider them here.

Concerning the possible quantum numbers of Higgses, 
they will be discussed in
in Section 3 where they are important e.g. in order to determine
whether or not all Yukawa couplings in D-brane scenarios are allowed.

Let us now try to determine the type I string scale $M_I$,
using the above information. On the one hand, from
(\ref{hypercharge}) one obtains the following relation at $M_I$:
\bea
\frac{1}{\alpha_Y(M_I)} = 
\frac{2}{\alpha_1(M_I)} + \frac{4c_2^2}{\alpha_2(M_I)} 
+ \frac{6c_3^2}{\alpha_3(M_I)}\ .
\label{couplings}
\eea
On the other hand, 
the usual RGE's for gauge couplings are given
by
\bea
\frac{1}{\alpha_j(M_I)}=\frac{1}{\alpha_j (M_Z)} 
+ \frac{b_j^{ns}}{2\pi}\ln\frac{M_{s}}{M_Z}
+ \frac{b_j^{s}}{2\pi}\ln\frac{M_I}{M_s}
\ ,
\label{running}
\eea
where $b_j^s$ ($b_j^{ns}$) with $j=2,3,Y$ are the coefficients of 
the supersymmetric (non-supersym\-metric) $\beta$-functions, and 
the scale $M_{s}$ corresponds to the supersymmetric threshold, 
200 GeV $\lsim M_s\lsim$ 1000 GeV. Thus
using (\ref{couplings}), (\ref{running}) and the fact that always
$c_2^2=1/4$
one can compute $M_I$ with the result
\bea
\ln\frac{M_I}{M_s}=
\frac{2\pi\left(\frac{1}{\alpha_Y (M_Z)}-\frac{2}{\alpha_1 (M_I)}-
\frac{1}{\alpha_2 (M_Z)}-
\frac{6c_3^2}{\alpha_3 (M_Z)}\right)
+\left(b_Y^{ns}-b_2^{ns}-6c_3^2 b_3^{ns}\right)
\ln\frac{M_s}{M_Z}
}
{\left(6c_3^2 b_3^s + b_2^s -b_Y^s\right)}
\ .
\label{MI}
\eea
%
%

Using the experimental values \cite{pdg} 
$M_Z=91.1870$ GeV, $\alpha_3 (M_Z)=0.1184$, 
$\alpha_2 (M_Z)=0.0338$, $\alpha_Y (M_Z)=0.01016$,
and the matter content of the minimal supersymmetric standard model
(MSSM), i.e. $b_3^s=3$, $b_2^s=-1$, $b_Y^s=-11$ and 
$b_3^{ns}=7$, $b_2^{ns}=19/6$, $b_Y^{ns}=-41/6$, one obtains for $c_3=-1/3$
\bea
\ln\frac{M_I}{M_s} = 33.09-\frac{1.05}{\alpha_1(M_I)}-1.22
\ln\frac{M_s}{M_Z}
\ .
\label{MI19}
\eea
For example, choosing the value of the coupling associated 
to the D$p_1$-brane in the range $0.07\lsim\alpha_1 (M_I)\lsim 0.1$ 
one obtains
$M_I\approx 10^{10-12}$ GeV. This scenario is shown in Fig.~2 
for $\alpha_1 (M_I)=0.1$ and $M_s=1$ TeV.
\begin{figure}[htb]
\begin{center}
\begin{tabular}{c}
\epsfig{file= 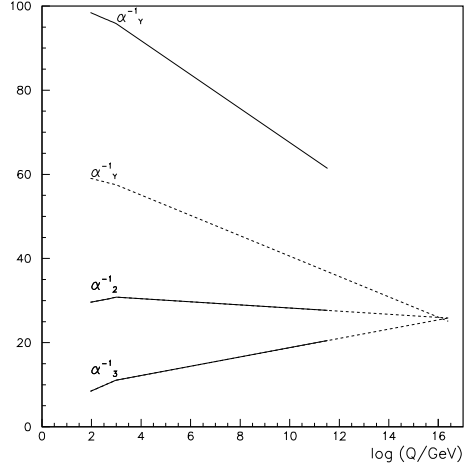, width=7cm}\\
\end{tabular}
\end{center} 
\caption{Running of the gauge couplings of the MSSM with energy $Q$ 
embedding the gauge groups within different sets of D$p$-branes (solid lines).
Due to the D-brane origin of the $U(1)$ gauge groups, relation
(\ref{couplings}) must be fulfilled. 
For comparison
the running of the MSSM couplings with
the usual normalization factor for the hypercharge, $3/5$,
is also shown with dashed lines.} 
\end{figure}
As discussed above, this "intermediate" initial scale is an
attractive possibility. 
Values of the coupling $\alpha_1 (M_I)$ 
smaller than 0.07 are not interesting since $M_I$
becomes also smaller, and therefore $m_{3/2}$ in (\ref{gravitino}) 
will be too low to be compatible with the experimental bounds on
supersymmetric particle masses.
Although the larger the coupling is, the larger $M_I$ becomes
(e.g. for $\alpha_1 (M_I)= 1$ one is even able to obtain  
$M_I\approx 5\times 10^{15}$ GeV),
one should be careful with the range of validity of the
perturbative regime.
%

On the other hand, the case $c_3=2/3$ is less interesting
since one obtains the upper bound 
$M_I\approx 3\times 10^{8}$ GeV.
%

It is worth noticing that non-supersymmetric scenarios can also be
analyzed with the above formula (\ref{MI}) with the substitutions
$M_s\rightarrow M_Z$, $b_i^s\rightarrow b_i^{ns}$.
For example, $M_I\approx 1$ TeV can be obtained with $\alpha_1 (M_I)\approx 
0.035$ for $c_3=-1/3$, and 
$\alpha_1 (M_I)\approx 
0.056$ for $c_3=2/3$.

\subsubsection{Scenarios with D$p_1$=D$p_3$ or D$p_1$=D$p_2$}

Let us now simplify the above analysis assuming that the D-brane
associated
to the $U(1)_1$ is on top of one of the other D-branes. In this case
we have two possibilities, either 
D$p_1$=D$p_3$ or D$p_1$=D$p_2$.
Let us start analyzing the possibility D$p_1$=D$p_3$, which
implies 
$\alpha_1=\alpha_3$. Then eq.(\ref{MI}) is still valid
with the substitutions 
$\frac{2}{\alpha_1 (M_I)}+
\frac{6c_3^2}{\alpha_3 (M_Z)}\rightarrow \frac{2+6c_3^2}{\alpha_3 (M_Z)}$,
$6c_3^2 b_3^{s,ns}\rightarrow (2+6c_3^2) b_3^{s,ns}$.
As a consequence, for $c_3=-1/3$, one obtains the following prediction:
$M_I\approx 6\times 10^{8}$ GeV, with $M_s=200$ GeV.
A slightly low value to be able to obtain $m_{3/2}\approx M_W$, as
discussed below eq.(\ref{gravitino}).
Obviously, the larger $M_s$ is, 
the smaller $M_I$ becomes.  
The case $c_3=2/3$ is much worse since $M_I\approx 100$ TeV.

The other scenario D$p_1$=D$p_2$, which implies 
$\alpha_1=\alpha_2$, does not improve the above situation.
One can use again eq.(\ref{MI}), but now with 
the substitutions
$\frac{2}{\alpha_1 (M_I)}+\frac{1}{\alpha_2 (M_Z)}
\rightarrow \frac{3}{\alpha_2 (M_Z)}$,
$b_2^{s,ns}\rightarrow 3b_2^{s,ns}$.
In particular, for $c_3=-1/3$,
$M_I\approx 500$ GeV, with $M_s=200$ GeV,
whereas $\ln \frac{M_I}{M_s}$ is even negative for $c_3=2/3$.

On the other hand, extra particles appear quite frecuently in
superstring theories. Since their presence will modify the
denominator in (\ref{MI}), one might obtain larger values for
$M_I$. For example, for D$p_1$=D$p_3$,
and restricting ourselves to the case of singlets,
$SU(2)$ doublets and colour triplets, one has 
\bea
(2+6c_3^2) b_3^s + b_2^s -b_Y^s= 18 -\frac{1}{2}(2+6c_3^2) n_3
-\frac{1}{2} n_2+q
\ ,
\label{extra}
\eea
where $q=\sum_{i=1}^{n_1} Y_i^2+2\sum_{j=1}^{n_2} Y_J^2+3\sum_{k=1}^{n_3}
Y_k^2$ and
$n_{1,2,3}$ is the number of extra singlets, doublets and
triplets that the model under consideration has. Extra $(3,2)$ 
representations under $SU(3)\times SU(2)$ must be introduced
in the formula for $q$ just as two triplets each.
For instance, assuming the presence of two copies of $d^c$+${\bar d}^c$ 
for the case $c_3=-1/3$,
one obtains $M_I\approx 4\times 10^{10}$ GeV 
($\approx 8\times 10^{9}$ GeV), with $M_s=200$ GeV (1 TeV).
Concerning the running of the couplings, this scenario is similar
to the one shown in Fig.~2.

As above, we can also
analyze non-supersymmetric scenarios.
A string scale of order a few TeV can be obtained without extra particles.
In particular, for $\alpha_1=\alpha_2$, $c_3=-1/3$
and $\alpha_1=\alpha_3$, $c_3=2/3$
we recover the results of \cite{Antoniadis}, 
$M_I\approx 300$ GeV and $M_I\approx 7$ TeV, 
respectively.

\subsubsection{Scenario without D$p_1$-brane}

Let us finally consider the scenario where
the $U(1)_Y$ is only a linear combination 
of the two $U(1)$ gauge groups arising from $U(3)_c$ and $U(2)_L$ 
within D$p_3$- and D$p_2$-branes respectively \cite{Rigolin2}.

As discussed in \cite{Antoniadis}, 
there is only one assignment of quantum numbers for quarks and leptons, 
in order to obtain the 
hypercharge of the standard model. The latter is given by
eq.~(\ref{hypercharge}) with $c_3=-1/3$, $c_2=-1/2$, $Q_1=0$, i.e. 
$Y=-\frac{1}{3}\sqrt 6 Q_3 -\frac{1}{2}\sqrt 4 Q_2$.
Whereas the charges $Q_3$ and $Q_2$ for $Q_u$, $d^c$ and
$L_e$ are as in the first assignment given in Table~1,
$Q_3=2/\sqrt{6},0$ and $Q_2=0,-2/\sqrt{4}$ for $u^c$, $e^c$.
Clearly, $u^c$ and $e^c$ must arise from open strings with
both ends on D$p_3$-branes and D$p_2$-branes, respectively.
As mentioned above, this is possible since these particles
can be obtained as the antisymmetric product
of two triplets of $SU(3)$ and doublets of $SU(2)$, respectively.

With the above hypercharge, instead of eq.~(\ref{couplings})
one obtains $\frac{1}{\alpha_Y(M_I)} = \frac{1}{\alpha_2(M_I)} 
+ \frac{2/3}{\alpha_3(M_I)}$,
and therefore eq.~(\ref{MI}) is still valid with $c_3=-1/3$ and 
the substitution 
$\frac{2}{\alpha_1 (M_I)}\rightarrow 0$.
As a consequence one can predict the string scale.
For example, for $M_s=200$ GeV ($M_s=1$ TeV)
one obtains
$M_I\approx 1.8\times 10^{16}$ GeV ($M_I\approx 10^{16}$ GeV). 
On the other hand, a non-supersymmetric scenario \cite{Antoniadis} gives rise
to a string scale which is too large, $M_I\approx 5\times 10^{13}$ GeV.

It is worth noticing \cite{Ibanez} that for $\alpha_3=\alpha_2$ one obtains
the standard GUT normalization for couplings
$\alpha_Y=\frac{3}{5}\alpha_2$,
and therefore $M_I\approx 2\times 10^{16}$ GeV.

\vspace{0.5cm}

\noindent We thus conclude that, concerning the string scale $M_I$,
the generic models analyzed above are very interesting
from the point of view of their predictivity.
Besides, the values obtained for $M_I$ can be 
accomodated in type I strings, choosing the
appropriate values of the moduli.
For instance, for the example studied below eq.~(\ref{extra}) 
the experimental values of couplings are obtained with
$M_I\approx 8\times 10^{9}$ GeV for the case $M_s=1$ TeV, 
and therefore the ratio
$\frac{\alpha_3 (M_I)}{\alpha_2 (M_I)}\approx 2$.
Let us assume that $SU(3)_c$ is embedded inside D9-branes 
and $SU(2)_L$ inside D$5_1$-branes. Then one has the following
relationships 
\bea
\frac{M_1M_2M_3}{M_I^2}= \frac{\alpha_3 M_{Planck}}{\sqrt 2}\  \ , \ \ \
\frac{M_1 M_I^2}{M_2M_3}= \frac{\alpha_{2} M_{Planck}}{\sqrt 2}\ ,
\label{relations}
\eea
where $M_i$, $i=1,2,3$, are the compactification masses associated to the 
compact radii $R_i$. Choosing $\frac{M_I^4}{M_2^2 M_3^2}\approx 1/2$ 
one is able to reproduce the above ratio.


\subsection{Embedding all gauge groups within the same set of D$p$-branes}

The fact that to obtain 
three copies of quark and leptons is difficult, 
when gauge groups come from different sets
of D$p_N$-branes, as mentioned above,
is one of the motivations in \cite{Ibanez} 
to embed all gauge interactions 
in the same set of D$p_N$-branes
($p_3=p_2=p_1$ in the notation above).

Here we will briefly review the results of 
Aldazabal, Ib\'a\~nez, Quevedo and 
Uranga \cite{Ibanez} concerning this issue. They are 
able to build 
$Z_3$ orientifold models with the gauge group
$SU(3)_c\times SU(2)_L\times U(1)_3\times U(1)_2\times U(1)_1$
embedded in D3-branes,
with no additional non-abelian factors.
They also argue that in the $Z_3$ orientifold,
which leads naturally to three
families, only the combination 
\bea
Y=-\frac{1}{3}\sqrt 6 Q_3-\frac{1}{2}\sqrt 4 Q_2+\sqrt 2 Q_1
\label{hypercharge2}
\eea
will be non-anomalous. It is worth noticing that this is 
precisely the hypercharge given in (\ref{hypercharge})
with $c_3=-1/3$ and $c_2=-1/2$, i.e. the first assignment of Table~1.
Likewise Fig.~1 with D$p_3$=D$p_2$=D$p_1$=D3, and all D3-branes
on top of each other, is 
also valid as a schematic representation of this type of
models. D$q$-branes in the figure are now D7-branes, which must
be introduced in order to cancel non-vanishing tadpoles.
Since $\alpha_3=\alpha_2=\alpha_1=\alpha$, 
instead of (\ref{couplings}) one obtains
\bea
\frac{1}{\alpha_Y (M_I)} = 
\frac{11/3}{\alpha (M_I)}\ ,
\label{couplings2}
\eea
which is not the standard GUT normalization for couplings. This is due
to the D-brane origin of the $U(1)$ gauge groups.

A model with all these properties was explicitly built in \cite{Ibanez}.
Although extra $U(1)'s$ on the D7-branes are present, they are
anomalous and therefore the associated gauge bosons have masses 
of the order of $M_I$.
In addition to D7-branes, anti-D7-branes trapped at different 
$Z_3$ fixed points are also present. Since they break supersymmetry
at the string scale $M_I$, they can be used as the hidden sector 
of supergravity theories.
Thus
this is an example of gravity mediated
supersymmetry breaking.

On the other hand, in this model not only quarks and leptons come in 
three generations but also Higgses, i.e. it contains 
two pairs of extra doublets with respect to the MSSM. 
In addition, three pairs of extra
colour triplets are also present. Unfortunately, this matter content
cannot give rise to the correct values for $\alpha_j (M_Z)$.
Although generically the extra triplets will be heavy, this does not
modify the previous result. One cannot exclude, however, the
possibility
that other models with the necessary matter content, in order to reproduce
the experimental values of couplings, might be built.
For example, 
if besides the matter content of the MSSM, 
we have six copies of $H_1$+$H_2$ and two copies of 
$d^c$+${\bar d}^c$ 
unification at around $M_I=10^{10}$ GeV, with
$\alpha (M_I)\approx 1/14$, is obtained. This scenario is
shown in Fig.~3 for $M_s=1$ TeV. 
\begin{figure}[htb]
\begin{center}
\begin{tabular}{c}
\epsfig{file= 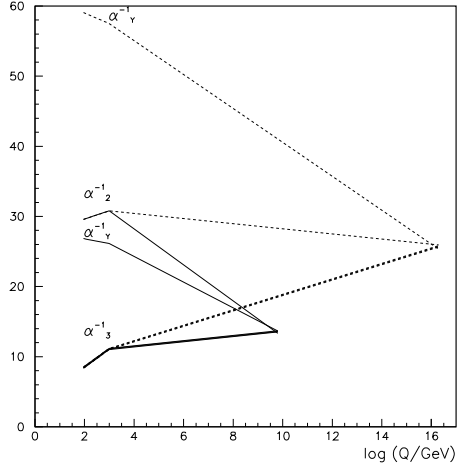, width=7cm}\\
\end{tabular}
\end{center} 
\caption{Running of the gauge couplings with energy $Q$
embedding all gauge groups within the same set of D3-branes (solid lines).
In addition to the matter content of the MSSM, extra Higgs doublets
and vector-like states
are also present. 
Due to the D-brane origin of the $U(1)$ gauge groups, the normalization
factor of the hypercharge is $3/11$ (see eq.~(\ref{couplings2})).
For comparison the running of the MSSM couplings with
the usual normalization factor for the hypercharge, $3/5$, is also
shown with dashed lines.} 
\end{figure}

It is worth noticing that
these values can be accomodated in type I strings, choosing the
appropriate values of the moduli. For example, with an isotropic
compact space, the string  
scale is given by:
\bea
M_I^4= \frac{\alpha M_{Planck}}{\sqrt 2} M_c^3\ ,
\label{gravitino2}
\eea
where $M_c$ is the compactification scale. 
Thus one gets $M_I\approx 10^{10-12}$ GeV with $M_c\approx 10^{8-10}$ GeV.

Let us finally mention that another model with the gauge group of the
standard model and three families has recently been built \cite{Bailin}.
The presence of additional Higgs doublets and
vector-like states allows an unification
scale at an intermediate value.

\section{Soft terms and Yukawa couplings in D-brane scenarios}

General formulas for the 
soft supersymmetry-breaking terms in D-brane constructions
were obtained in \cite{Rigolin2}, under the assumption
of dilaton/moduli supersymmetry breaking 
\cite{BIM22}-\cite{BIM},
using the parametrization introduced in \cite{BIM}.
On the other hand,
general Yukawa couplings in D-brane constructions have 
been studied in \cite{Yukawas,Rigolin2}.
Since we need to use these results to obtain the
soft terms and Yukawa couplings 
associated to the D-brane scenarios discussed above,
we summarize them in the Appendix A.


\subsection{Embedding the gauge groups within different sets of D$p$-branes}

\subsubsection{General scenario with $Dp_3\neq Dp_2\neq Dp_1\neq Dp_3$}

For the sake of concreteness, let us assume the following distribution
of D-branes in the scenario proposed in Subsection 2.1.
D$p_3$-branes are D9-branes, 
D$p_1$-branes are D$5_3$-branes, 
D$p_2$-branes are D$5_1$-branes, 
and finally D$q$-branes are D$5_2$-branes.
Then,  
the first assignment of Table 1 shown schematically in Fig.~1
gives rise to the following soft masses,
using formulas (\ref{gaugino111}) and (\ref{scalar111}) in the
Appendix A. The gaugino masses are given
by:
%
%
\bea
M_3 & = & \sqrt{3} m_{3/2} \sin \theta \ , \nn \\
M_{2} & = & \sqrt{3}  m_{3/2}\ \Theta_1 \cos \theta  \ , \nn\\
M_{Y} & = &  \sqrt{3}  m_{3/2}\ \alpha_Y (M_I)
\left(\frac{2}{\alpha_1 (M_I)}\Theta_3 \cos \theta 
+\frac{1}{\alpha_2 (M_I)}\Theta_1 \cos \theta
+\frac{6c_3^2}{\alpha_3 (M_I)}\sin \theta
\right)
\label{gaugino1}
\eea
where it is worth noticing that relation (\ref{couplings}) has been
taken into account in order to obtain the gaugino mass associated to
the gauge group $U(1)_Y$.
The scalar masses are given by:
\bea
m^2_{Q_u} & = & m_{3/2}^2\left[1 - 
\frac{3}{2}  \left(1 - \Theta_{1}^2 \right)
\cos^2 \theta \right] \ , \nn \\
m^2_{d^c} & = & m_{3/2}^2\left[1 - 
\frac{3}{2}  \left(1 - \Theta_{2}^2 \right)
\cos^2 \theta \right] \ , \nn \\
m^2_{u^c} & = & m_{3/2}^2\left[1 - 
\frac{3}{2}  \left(1 - \Theta_{3}^2 \right)
\cos^2 \theta \right] \ , \nn \\
m^2_{e^c} & = & m_{3/2}^2\left[1- \frac{3}{2} 
\left(\sin^2\theta + \Theta_{1}^2 \cos^2\theta  \right)\right] \ , \nn \\
m^2_{L_e} & = & m_{3/2}^2\left[1- \frac{3}{2} 
\left(\sin^2\theta + \Theta_{3}^2 \cos^2\theta  \right)\right] \ .
\label{scalars1}
\eea
%
%
Note that quarks of type $Q_u$, $d^c$ and $u^c$
are states $C^{95_1}$, $C^{95_2}$ and $C^{95_3}$ respectively,
whereas leptons of type $e^c$ and $L_e$ are states
$C^{5_3 5_2}$ and $C^{5_1 5_2}$ respectively.

These soft terms (\ref{gaugino1}) and (\ref{scalars1}) are generically
non-universal.
For example in the 
overall modulus limit ($\Theta_{1,2,3}=1/\sqrt 3$) 
universality cannot be obtained. The dilaton limit
($\sin^2\theta=1$)
would give rise to tachyonic states $e^c$ and $L_e$.

Obviously, the other three assignments of quantum numbers in Table 1
give rise to the same gaugino masses (\ref{gaugino1}).
The differences arise for some of the soft scalar masses in (\ref{scalars1}).
For the second assignment the masses of leptons of type $L_e$
in (\ref{scalars1}) must be replaced by
\bea
m^2_{L_e} &  = & m_{3/2}^2\left[1- \frac{3}{2} 
\left(\sin^2\theta + \Theta_{2}^2 \cos^2\theta  \right)\right] \ . 
\label{leptons}
\eea
For the third assignment the masses of quarks of type $u^c$ and $d^c$
must be exchanged in (\ref{scalars1}), i.e.
\bea
m^2_{u^c} & = & m_{3/2}^2\left[1 - 
\frac{3}{2}  \left(1 - \Theta_{2}^2 \right)
\cos^2 \theta \right] \ , \nn \\
m^2_{d^c} & = & m_{3/2}^2\left[1 - 
\frac{3}{2}  \left(1 - \Theta_{3}^2 \right)
\cos^2 \theta \right] \ . 
\label{sud}
\eea
Finally, for the fourth assignment, both modifications
(\ref{leptons}) and (\ref{sud}) must be included in eq.(\ref{scalars1}).

Concerning the soft Higgs masses, we need to know 
the quantum numbers $Q_{3,2,1}$ of the two
Higgs doublets of the supersymmetric standard model.
For the first and third assignments of Table 1 whose hypercharges are
$Y=-\frac{1}{3}\sqrt 6 Q_3-\frac{1}{2}\sqrt 4 Q_2+\sqrt 2 Q_1$
and $Y=\frac{2}{3}\sqrt 6 Q_3-\frac{1}{2}\sqrt 4 Q_2+\sqrt 2 Q_1$ 
respectively, there are two possible assignments for the
Higgs with hypercharge $1/2$, $H_2(0, \frac{1}{\sqrt 4}, \frac{1}{\sqrt 2})$ 
and $H_2(0, -\frac{1}{\sqrt 4}, 0)$.
For the Higgs with hypercharge $-1/2$, there are also 
two possible assignments 
$H_1(0, \frac{1}{\sqrt 4}, 0)$ and
$H_1(0, -\frac{1}{\sqrt 4}, -\frac{1}{\sqrt 2})$. 
Thus we have four possible combinations:
\bea
H_2(0, 1, 1)\  \ , \ \ \
H_1(0, 1, 0)\ 
\label{assig1}
\\
H_2(0, 1, 1)\  \ , \ \ \
H_1(0, -1, -1)\ 
\label{assig2}
\\
H_2(0, -1, 0)\  \ , \ \ \
H_1(0, -1, -1)\ 
\label{assig3}
\\
H_2(0, -1, 0)\  \ , \ \ \
H_1(0, 1, 0)\ 
\label{assig4}
\eea
where for simplicity we have multiplied the quantum numbers by
$\sqrt{2N}$
as in Table 1.
For example, combination (\ref{assig1}) is the one shown in Fig.~1,
where $H_2$ is a state $C^{5_1 5_3}$ and $H_1$ is a state
$C^{5_1 5_2}$.
The corresponding soft masses, using again formulas (\ref{scalar111}),
are given respectively by
\bea
&& \!\!\!\!\!\!\!\!\!\!\!\!\!\!\! m^2_{H_2}  =  m_{3/2}^2\left[1- \frac{3}{2} 
\left(\sin^2\theta + \Theta_{2}^2 \cos^2\theta  \right)\right] \ , 
m^2_{H_1}  =  m_{3/2}^2\left[1- \frac{3}{2} 
\left(\sin^2\theta + \Theta_{3}^2 \cos^2\theta  \right)\right] 
\label{Higgses1}
\\
&& \!\!\!\!\!\!\!\!\!\!\!\!\!\!\! m^2_{H_2}  =  m^2_{H_1} = m_{3/2}^2\left[1- \frac{3}{2} 
\left(\sin^2\theta + \Theta_{2}^2 \cos^2\theta  \right)\right]  
\label{Higgses2}
\\
&& \!\!\!\!\!\!\!\!\!\!\!\!\!\!\! m^2_{H_2}  =  m_{3/2}^2\left[1- \frac{3}{2} 
\left(\sin^2\theta + \Theta_{3}^2 \cos^2\theta  \right)\right] \ , 
m^2_{H_1}  =  m_{3/2}^2\left[1- \frac{3}{2} 
\left(\sin^2\theta + \Theta_{2}^2 \cos^2\theta  \right)\right] 
\label{Higgses3}
\\
&& \!\!\!\!\!\!\!\!\!\!\!\!\!\!\! m^2_{H_2} =  m^2_{H_1} = m_{3/2}^2\left[1- \frac{3}{2} 
\left(\sin^2\theta + \Theta_{3}^2 \cos^2\theta  \right)\right]\ .
\label{Higgses4}
\eea

With respect to the second and fourth assignment of quantum numbers
in Table 1, it is worth noticing that
their hypercharges are
equal to the ones of first and third assignment respectively,
but with an opposite sign in front of $Q_2$. As a consequence,
the four combinations of quantum numbers 
(\ref{assig1}-\ref{assig4})
are still valid using an opposite sign for the value of $Q_2$.
Thus the corresponding soft masses are like in
eqs. 
(\ref{Higgses1}-\ref{Higgses4}).

Summarizing, 
we have obtained in total 
sixteen scenarios with different soft terms.

Concerning the soft trilinear parameters, since these are related to Yukawa
couplings we need to discuss first the structure of the latter. 
This can be carried out straightforwardly 
taking into account the previous information
about
quantum numbers and
the formula for the renormalizable 
Yukawa Lagrangian (\ref{superpottt}) in the Appendix A.
The sixteen possible scenarios will have in principle 
different Yukawa couplings since fields
$u^c$, $d^c$, $L_e$, $H_2$ and $H_1$ are attached to different D-branes.

Let us 
concentrate on the eight scenarios which are more realistic from the
phenomenological point of view. Following the discussion of Subsection
2.1.1
these are the ones with $c_3=-1/3$, i.e. the first assignment of 
Table~1 with the four possible combinations for Higgses 
(\ref{assig1}-\ref{assig4}), and
the second assignment of Table~1 with the same combinations but with
an opposite sign for the value of $Q_2$ as discussed above.
Let us also assume that we have three copies of quarks and leptons.
For instance, the first assignment 
with Higgses as in (\ref{assig1}) implies that couplings
$\sum_{a,b}Y_u^{ab}H_2 Q_u^a u^c_b$
correspond to 
$C^{5_35_1}C^{95_1}C^{95_3}$,
couplings 
$\sum_{a,b}Y_d^{ab}H_1 Q_u^a d^c_b$
correspond to 
$C^{5_15_2}C^{95_1}C^{95_2}$,
and finally couplings 
$\sum_{a,b}Y_e^{ab}H_1 L_e^a e^c_b$
correspond to 
$C^{5_15_2}C^{5_15_2}C^{5_35_2}$, where $a, b$ are family indices.
Whereas the last type of couplings is forbidden,
as can be seen from\footnote{This is also obtained
realizing that the sum of the $U(1)$ charges of the fields
is non-vanishing.} (\ref{superpottt}) in the Appendix A,
the other two are allowed with the result
\begin{equation}
Y_{u,d}=g_{q,1}\ \hat Y
\ ,\ Y_e=0 \ ,
\label{Yukawas}
\end{equation}
where $\hat Y$ is defined as
\begin{equation}
\hat Y=\bmat{ccc}1 & 1 & 1\\ 1 & 1 & 1 \\
1 & 1 & 1 \emat \ ,
\label{Yukawos}
\end{equation}
$g_1$ is the gauge coupling associated to the $U(1)_1$ in the
D$5_3$-brane 
and $g_q$ is the gauge coupling associated to the D$5_2$-brane.
An analysis of the above ``democratic'' texture for $Y_{u,d}$
can be found in \cite{Fritzsch}. Lepton masses might appear from
non-renormalizable couplings.

The Yukawa couplings for the other three combinations of Higgses
(\ref{assig2}-\ref{assig4})
can be obtained straightforwardly as for the previous one, with 
the result
\begin{eqnarray}
Y_{u,e}=g_{q,3}\ \hat Y
\ ,\ Y_d=0 \ ,
\label{Yukawas2as}
\\
Y_{e}=g_{3}\ \hat Y
\ ,\ Y_{u,d}=0 \ ,
\label{Yukawas3as}
\\
Y_{d}=g_{1}\ \hat Y
\ ,\ Y_{u,e}=0 \ ,
\label{Yukawas4as}
\end{eqnarray}
respectively. Here 
$g_3$ is the gauge coupling associated to the $SU(3)_c$ in the
D9-brane. 

On the other hand,
for the second assignment of Table~1 the Yukawa couplings are given by
\begin{eqnarray}
Y_{u,d,e}=g_{q,1,3}\ \hat Y\ ,
\label{Yukawas5as}
\\
Y_{u}=g_{q}\ \hat Y
\ ,\ Y_{d,e}=0 \ ,
\label{Yukawas6as}
\\
Y_{u,d,e}=0 \ ,
\label{Yukawas7as}
\\
Y_{d,e}=g_{1,3}\ \hat Y
\ ,\ Y_{u}=0 \ .
\label{Yukawas8as}
\end{eqnarray}
It is worth noticing that in principle some of these scenarios
seem to be hopeless. For instance, it
is difficult to imagine how scenario with Yukawa couplings
(\ref{Yukawas3as}) may give rise to the 
observed fermion mass hierarchies with quark masses
arising from 
non-renormalizable terms.

Now that the structure of Yukawa couplings is known we can compute
the corresponding trilinear parameters using 
eq.(\ref{trilin1}). 
Obviously, when Yukawa couplings are vanishing
trilinear parameters are also vanishing. 
When Yukawa couplings are non-vanishing, the A terms acquire the 
following values:
\bea
A_{u} & = &  \frac{\sqrt 3}{2}m_{3/2}
   \left[\left(\Theta_{2} - \Theta_1  - \Theta _{3} \right) \cos\theta  
- \sin\theta \right] \ \hat Y\ ,
\label{trintrin}
\\
A_{d} & = &  \frac{\sqrt 3}{2}m_{3/2}
   \left[\left(\Theta_{3} - \Theta_1  - \Theta _{2} \right) \cos\theta  
- \sin\theta \right] \ \hat Y\ ,
\label{trintrintrin}
\\
A_{e} & = &  \frac{\sqrt 3}{2}m_{3/2}
   \left[\sin\theta - \left(\Theta_{1} + \Theta_2  + \Theta _{3}
     \right) 
\cos\theta  \right] \hat Y\ .
\label{trilin11}
\eea
For example, Yukawa couplings (\ref{Yukawas})
have associated A terms given by (\ref{trintrin}),
(\ref{trintrintrin}) and $A_e=0$,
Yukawa couplings (\ref{Yukawas2as}) have associated A-terms
(\ref{trintrin}), (\ref{trilin11}) and $A_d=0$, etc.

\subsubsection{Scenarios with D$p_1$=D$p_3$ or D$p_1$=D$p_2$}

As we discussed in Subsection 2.1.2,
scenarios with D$p_1$=D$p_3$ 
are more interesting from the phenomenological point of view than
scenarios with D$p_1$=D$p_2$, thus we 
will concentrate in the former. In any case, the analysis of the
other scenario can be carried out along similar lines.
The first attempts to study these scenarios and their phenomenology,
in particular CP phases, Yukawa textures, and dark matter,
were carried out in \cite{Kane}, \cite{Kane2}, and 
\cite{khalil,Nath2,Arnowitt,Bailindm}, respectively. However, they
consider in fact 
toy scenarios since 
the important issue of the D-brane origin of the $U(1)_Y$ gauge group
as a combination of other $U(1)$'s 
and its influence on the matter distribution (see e.g. Fig~1) 
was not included in their analyses.
Thus we will take into account the discussion of Section 2
concerning this issue 
in order to obtain
the soft terms and Yukawa
couplings, as we already did with the general scenario of
the previous subsection.

Assuming the same distribution of D-branes as in the previous
subsection, we have that
D$p_1$- and D$p_3$-branes are D9-branes, 
D$p_2$-branes are D$5_1$-branes, 
and finally D$q$-branes are D$5_2$-branes.
Then, the first assignment of Table 1 
gives rise to the following gaugino masses:
\bea
M_3 & = & \sqrt{3} m_{3/2} \sin \theta \ , \nn \\
M_{2} &  = & \sqrt{3}  m_{3/2}\ \Theta_1 \cos \theta  \ , \nn \\
M_{Y} & = &  \sqrt{3}  m_{3/2}\ \alpha_Y (M_I)
\left(\frac{1}{\alpha_2 (M_I)}\Theta_1 \cos \theta
+\frac{2+6c_3^2}{\alpha_3 (M_I)}\sin \theta
\right) ,
\label{gauginos1}
\eea
and scalar masses:
\bea
m^2_{Q_u} & = & m_{3/2}^2\left[1 - 
\frac{3}{2}  \left(1 - \Theta_{1}^2 \right)
\cos^2 \theta \right] \ , \nn \\
m^2_{d^c} & = & m_{3/2}^2\left[1 - 
\frac{3}{2}  \left(1 - \Theta_{2}^2 \right)
\cos^2 \theta \right] \ , \nn \\
m^2_{u^c_i} & = & m_{3/2}^2\left[1 - 
{3} \Theta_{i}^2 \cos^2\theta \right]\ , \nn \\
m^2_{e^c} & = & m_{3/2}^2\left[1 - 
\frac{3}{2}  \left(1 - \Theta_{2}^2 \right)
\cos^2 \theta \right] \ , \nn \\
m^2_{L_e} & = & m_{3/2}^2\left[1- \frac{3}{2} 
\left(\sin^2\theta + \Theta_{3}^2 \cos^2\theta  \right)\right] \ .
\label{scalarss1}
\eea
Here $i=1,2,3$ labels the three complex compact dimensions.
Thus $u^c_i$ are states $C^9_i$ coming from open strings 
starting and ending on D9-branes.
These fields behave quite similarly to untwisted sectors 
of perturbative heterotic orbifolds. It is then natural to 
use this index
as a family index. For example we will
take $u^c_1=t^c$, $u^c_2=c^c$ and $u^c_3=u^c$.
On the other hand, quarks of type $Q_u$, $d^c$
are states $C^{95_1}$ and $C^{95_2}$ respectively,
whereas leptons of type $e^c$ and $L_e$ are states
$C^{9 5_2}$ and $C^{5_1 5_2}$ respectively.
As discussed below eq.(\ref{scalars1}) 
these soft terms are also generically non-universal.
It is worth noticing here 
that due to the family index $i$ a
potential problem due to flavor-changing neutral currents (FCNC)
may arise. Being conservative\footnote{Recall that the relevant mass terms
are the low-energy ones, not those generated at the string scale.
As discussed e.g. in \cite{BIM}, 
one has to do the low-energy running of the scalar masses, and,
for the squark case, for gluino masses heavier than (or of the same
order as) the scalar masses, there are large flavor-independent gluino
loop contributions which are the dominant source of scalar masses.
}
we can avoid it by imposing
$\Theta_2=\Theta_3$.
This constraint will be used in Section 4 when discussing 
neutralino-proton cross sections in this scenario.

For the second assignment, the value of $m^2_{L_e}$ must be
replaced 
by
\bea
m^2_{L_e}=m_{3/2}^2\left[1 - 
\frac{3}{2}  \left(1 - \Theta_{1}^2 \right)
\cos^2 \theta \right] \ . 
\label{Higgsess12}
\eea
For the third assignment the masses of quarks of type $u^c$ and $d^c$
must be exchanged in (\ref{scalarss1}), i.e. 
\bea
m^2_{u^c} & = & m_{3/2}^2\left[1 - 
\frac{3}{2}  \left(1 - \Theta_{2}^2 \right)
\cos^2 \theta \right] \ , \nn \\
m^2_{d^c_i} & = & m_{3/2}^2\left[1 - 
{3} \Theta_{i}^2 \cos^2\theta \right]
\ . 
\label{Higgsess123}
\eea
For the fourth assignment, both modifications (\ref{Higgsess12}) and 
(\ref{Higgsess123}) must be
included
in eq.(\ref{scalarss1}).

Finally, the Higgs masses corresponding to the four combinations
obtained in eqs. 
(\ref{assig1}-\ref{assig4}) are
\bea
&& \!\!\!\!\!\!\!\!\!\!\!\!\!\!\! m^2_{H_2}  =   m_{3/2}^2\left[1 - 
\frac{3}{2}  \left(1 - \Theta_{1}^2 \right)
\cos^2 \theta \right] \ ,
m^2_{H_1}  =  m_{3/2}^2\left[1- \frac{3}{2} 
\left(\sin^2\theta + \Theta_{3}^2 \cos^2\theta  \right)\right] \ ,
\label{Higgsess1}
\\
&& \!\!\!\!\!\!\!\!\!\!\!\!\!\!\! m^2_{H_2}  = m^2_{H_1} = m_{3/2}^2\left[1 - 
\frac{3}{2}  \left(1 - \Theta_{1}^2 \right)
\cos^2 \theta \right] \ ,
\label{Higgsess2}
\\
&& \!\!\!\!\!\!\!\!\!\!\!\!\!\!\! m^2_{H_2}  =  m_{3/2}^2\left[1- \frac{3}{2} 
\left(\sin^2\theta + \Theta_{3}^2 \cos^2\theta  \right)\right]\ ,
m^2_{H_1}  =  m_{3/2}^2\left[1 - 
\frac{3}{2}  \left(1 - \Theta_{1}^2 \right)
\cos^2 \theta \right] \ ,
\label{Higgsess3}
\\
&& \!\!\!\!\!\!\!\!\!\!\!\!\!\!\! m^2_{H_2}  = m^2_{H_1} = m_{3/2}^2\left[1- \frac{3}{2} 
\left(\sin^2\theta + \Theta_{3}^2 \cos^2\theta  \right)\right] \ , 
\label{Higgsess4}
\eea
respectively. For example $H_2$, $H_1$ in (\ref{Higgsess1}) 
are states $C^{95_1}$, $C^{5_15_2}$.

The analysis of Yukawa couplings and trilinear parameters
can be carried out as in the previous subsection,
assuming again three copies of quark and leptons 
$Q_u, d^c, L_e, e^c$.
As an example let us consider the second assignment of Table~1
with Higgses as in (\ref{assig2}) with an opposite sign for $Q_2$.
From eq.(\ref{superpottt}) 
we deduce that the only allowed type of coupling, which
corresponds 
to $C^{95_1}C^{95_1}C^{9}_1$, is
$H_2 Q_u^a t^c$. The result for Yukawa couplings is then
\begin{equation}
Y_{u}=g_{3}\ \tilde Y\ ,\ Y_{d,e}=0\ ,
\label{Yukawasss}
\end{equation}
where $\tilde Y$ is defined as 
\begin{equation}
\tilde Y=\bmat{ccc}0 & 0 & 1\\ 0 & 0 & 1 \\
0 & 0 & 1 \emat \ .
\label{Yukawosss}
\end{equation}
This structure for Yukawa matrices and its viability
has been studied in \cite{Kane2}.  
Other results for Yukawa couplings arise for the other interesting scenarios.
For example for the first assignment of Table~1
with Higgses as in
(\ref{assig1}), whereas $Y_u$ and $Y_e$ are still as above,
$Y_d$ has a ``democratic'' 
matrix structure as in
(\ref{Yukawos}). This structure was also analyzed in \cite{Kane2}. 

Concerning the 
trilinear parameters, these can be computed using again
(\ref{trilin1}). For instance, Yukawa couplings (\ref{Yukawasss})
have associated
\bea
A_{u} & = & - {\sqrt 3}m_{3/2} 
\sin\theta\ \tilde Y
\ ,\ A_{d,e}=0 \ . 
\label{trilinss1}
\eea
%


\subsubsection{Scenario without D$p_1$-brane}

In this scenario the gauge groups of the standard model
arise only from
D$p_3$-branes, which are D9-branes, and 
D$p_2$-branes, which are D$5_1$-branes. 
Then, following the discussion in Subsection 2.1.3,
quarks of type $Q_u$ are states $C^{95_1}$,
quarks of type $d^c$
are states $C^{95_2}$, and 
leptons of type $L_e$ are states $C^{5_1 5_2}$.
On the other hand, quarks of type $u^c$ are states $C^{9}_i$
and leptons of type $e^c$ are states $C^{5_1}_i$.
As mentioned before, it is natural to use this index $i$ as a family
index.
The only combination of quantum numbers which is now allowed for Higgses
is (\ref{assig4}), and therefore $H_1$, $H_2$ are states $C^{5_1 5_2}$.

Using again eqs. (\ref{gaugino111}) and (\ref{scalar111}) 
we can obtain the soft terms for this scenario.
In particular, the gaugino masses are
\bea
M_3 & = & \sqrt{3} m_{3/2} \sin \theta \ , \nn \\
M_{2} & = & \sqrt{3}  m_{3/2}\ \Theta_1 \cos \theta  \ , \nn\\
M_{Y} & = &  \sqrt{3}  m_{3/2}\ \alpha_Y (M_I)
\left(\frac{1}{\alpha_2 (M_I)}\Theta_1 \cos \theta
+\frac{2/3}{\alpha_3 (M_I)}\sin \theta
\right) ,
\label{gaugino0}
\eea
and the scalar masses are
\bea
m^2_{Q_u} &  = &  m_{3/2}^2\left[1 - 
\frac{3}{2}  \left(1 - \Theta_{1}^2 \right)
\cos^2 \theta \right] \ , \nn \\
m^2_{d^c} & = & m_{3/2}^2\left[1 - 
\frac{3}{2}  \left(1 - \Theta_{2}^2 \right)
\cos^2 \theta \right] \ , \nn \\
m^2_{L_e} & = &  m^2_{H_2}\ =\ m^2_{H_1}\  =\  m_{3/2}^2
\left[1- \frac{3}{2} 
\left(\sin^2\theta + \Theta_{3}^2 \cos^2\theta  \right)\right] \ , \nn \\
m^2_{u^c_i} & = & m_{3/2}^2\left[1 - 
{3} \Theta_{i}^2 \cos^2\theta \right]\ , \nn \\
m^2_{e^c} & = & m_{3/2}^2\left[1 - 
{3} \sin^2\theta \right]\ , \nn \\
m^2_{\mu^c} & = & \ m_{3/2}^2\left[1- 3 \Theta_{3}^2 
\cos^2\theta \right] \ , \nn \\
m^2_{\tau^c} & = & \ m_{3/2}^2\left[1- 3 \Theta_{2}^2 
\cos^2\theta \right] \ ,
\label{scalars0}
\eea
where 
the following assignments have been 
used for the leptons.
$e^c$ is a state $C^{5_1}_1$, 
$\mu^c$ is a state $C^{5_1}_2$, and finally 
$\tau^c$ is a state $C^{5_1}_3$.
As in the previous two scenarios these soft terms
are also generically non-universal.

Concerning Yukawa couplings,
these are allowed
for leptons
in this scenario since $C^{5_1 5_2}C^{5_1 5_2}C^{5_1}_3$ exists.
Assuming three copies of leptons $L_e$, one obtains
\begin{equation}
Y_{e}=g_{2}\ \tilde Y
\ ,
\label{Yukawasss0}
\end{equation}
where $\tilde Y$ has been defined in (\ref{Yukawosss}) and
$g_2$ is the gauge coupling associated to the $SU(2)_L$ in the
D$5_1$-brane. 
The corresponding trilinear parameters are
\bea
A_{e} & = & - {\sqrt 3}m_{3/2}\ \Theta_1
\cos\theta\ \tilde Y
\ . 
\label{trilinss1}
\eea
Notice however that the above assignment for leptons may be
problematic concerning FCNC since 
generically $m^2_{e^c}\neq  m^2_{\mu^c}$. We can avoid that
potential problem choosing $\tau^c$ as a state $C^{5_1}_1$ and e.g.
$e^c$ as a state $C^{5_1}_2$,  
$\mu^c$ as a state $C^{5_1}_3$. Then we have to choose  
$u^c_1=t^c$.
Now, imposing $\Theta_2=\Theta_3$, FCNC will not be present.   
This constraint will be used in Section 4 when discussing
neutralino-proton
cross sections in this scenario. Instead of the matrix structure
(\ref{Yukawasss0}) there will be a new matrix
with the non-vanishing entries in the
second column.

On the other hand, Yukawa couplings for quarks of type $u$ are
vanishing since couplings $C^{5_1 5_2}C^{9 5_1}C^{9}_i$ are forbidden.
However, couplings $C^{5_1 5_2}C^{9 5_1}C^{9 5_2}$ are allowed
and therefore Yukawa couplings for quarks of type $d$ exist.
The matrix structure is like  
%
%
in (\ref{Yukawos}).
In any case, as discussed for other scenarios in subsection 3.1.1,
is difficult to imagine how this scenario may give rise to the 
observed fermion mass hierarchies with masses of quarks of type $u$
arising from 
non-renormalizable terms.

\subsection{Embedding all gauge groups within the same set of D$p$-branes}

As discussed in Subsection 2.2,
the D-brane model constructed in \cite{Ibanez},
where all gauge groups are embedded in 3-branes,
is very interesting.
We will analyze here the soft terms and Yukawa couplings of the model.
 
Let us recall that 
D$p_{3,2,1}$-branes are in this scenario D$3$-branes, 
and D$q$-branes are D$7_{1,2,3}$-branes.
The distribution of matter is like in Fig.~1 with all D$3$-branes
on top of each other.
Taking into account that 
under a $T$-duality transformation with respect to the three 
compact dimensions the 9-branes transform into 3-branes and the
$5_i$-branes into $7_i$-branes, still the formulas for 
soft terms and Yukawas  
will be identical to the ones in the Appendix A,
(\ref{gaugino111}-\ref{superpottt}), 
with the obvious replacements 
$9\rightarrow 3$, $5_i\rightarrow 7_i$ everywhere. 
This implies the following gaugino masses:
\bea
M_3  = M_2 = M_Y = \sqrt{3} m_{3/2} \sin \theta\ ,
\label{gaugino3}
\eea
and scalar masses:
\bea
m^2_{Q_u^i}  =  m^2_{u^c_i} = m^2_{H_2^i} =
m_{3/2}^2\left(1 - {3} \Theta_{i}^2 
\cos^2 \theta \right) \ , \nn \\
m^2_{d^c_i}  = m^2_{L_e^i} = m^2_{e^c_i} = m^2_{H_1^i} =
m_{3/2}^2\left[1 - 
\frac{3}{2}  \left(1 - \Theta_{i}^2 \right)
\cos^2 \theta \right] \ .
\label{scalars3}
\eea
Note that  $Q_u$, $u^c$, and $H_2$ 
are states $C^{3}_i$,
whereas $d^c$, $e^c$, $L_e$ and $H_1$ are states
$C^{3 7_i}$. Thus due to the index $i=1,2,3$ 
three families arise in this model in a natural way.
For example we will take
$u^c_1=u^c$, $u^c_2=c^c$, $u^c_3=t^c$, etc.
In order to avoid FCNC we may impose $\Theta_1=\Theta_2$.
It is also worth noticing that universality can be obtained 
in the dilaton 
($\sin^2\theta=1$)
and
overall modulus ($\Theta_{1,2,3}=1/\sqrt 3$) 
limits unlike the scenarios in Subsection 3.1. 

Let us now analyze the Yukawa couplings of the model \cite{Ibanez}.
Couplings of the type $C^3_1 C^3_2 C^3_3$ are allowed.
Assuming that the physical Higgs $H_2$ corresponds to
$C^3_1$, the following couplings exist:
$g H_2 Q_c t^c$ and $g H_2 Q_t c^c$, i.e.
\begin{equation}
Y_{u}=g\ \bar Y
\ ,
\label{mas}
\end{equation}
where $\bar Y$ is defined as
\begin{equation}
\bar Y=
\bmat{ccc}0 & 0 & 0\\ 0 & 0 & 1 \\
0 & 1 & 0 \emat 
\ .
\label{mass}
\end{equation}
The corresponding trilinear parameters
are 
%
\bea
A_{u} & = & - {\sqrt 3}m_{3/2} 
\sin\theta\ \bar Y
\ . 
\label{masss}
\eea
Let us remark that in the presence of discrete torsion
$g H_2 Q_t t^c$ may also be present \cite{Ibanez}.
On the other hand, 
couplings of the type $C^3_i C^{3 7_i} C^{3 7_i}$ 
are also allowed. Assuming that the physical Higgs
$H_1$ corresponds to $C^{3 7_3}$,
the coupling $g Q_t b^c H_1$ also exists, i.e.
\begin{equation}
Y_{b}=g\ \bar{\bar Y}
\ ,
\label{mes}
\end{equation}
where $\bar{\bar Y}$ is defined as
\begin{equation}
\bar{\bar Y} =
\bmat{ccc}0 & 0 & 0\\ 0 & 0 & 0 \\
0 & 0 & 1 \emat 
\ .
\label{mess}
\end{equation}
The trilinear parameters
are 
%
\bea
A_{b} & = & - {\sqrt 3}m_{3/2} 
\sin\theta\ \bar{\bar Y}
\ . 
\label{messs}
\eea
%
More involved couplings with off-diagonal entries in the
matrix for quarks of type $d$ are possible in some circumstances \cite{Ibanez}.
Finally, renormalizable Yukawa couplings for leptons are not present
since they are of the type $C^{37} C^{37} C^{37}$
and these are not allowed.

\section{Neutralino-nucleon cross sections in $D$-brane scenarios}

Recently 
there has been some theoretical activity 
\cite{Bottino}-\cite{gomez} 
analyzing the compatibility of regions
in the parameter space of 
the MSSM
with the sensitivity of current 
(DAMA \cite{experimento1}, CDMS \cite{experimento2}, 
HEIDELBERG-MOSCOW \cite {HM}, HDMS prototype \cite{HDMS}, UKDMC \cite{mine}, 
CANFRANC \cite{canfranc})
and projected 
(GENIUS \cite{GENIUS}, DAMA 250 kg. \cite{experimento1}, 
CDMS Soudan \cite{experimento2}, etc.)
dark matter detectors. In particular,
DAMA and CDMS
are sensitive to a 
neutralino-nucleon cross 
section
$\sigma_{\tilde\chi_1^0-p}$ 
in the range of $10^{-6}$--$10^{-5}$ pb.
Working in the supergravity framework for 
the MSSM with universal soft terms, 
it was pointed out 
in \cite{Bottino,arna,arnowitt,gomez} 
that the large $\tan\beta$ regime   
allows regions where the above mentioned range 
of $\sigma_{\tilde\chi_1^0-p}$ is reached.
Besides, working with non-universal soft scalar masses $m_{\alpha}$,
$\sigma_{\tilde\chi_1^0-p}\approx 10^{-6}$
pb was also found for small values of $\tan\beta$, if $m_{\alpha}$ fulfil 
some special conditions \cite{Bottino,arna,arnowitt}.
In particular, this was obtained for $\tan\beta\gsim 25$ 
($\tan\beta\gsim 4$) working with universal (non-universal) soft terms
in \cite{arnowitt}.
The case of non-universal gaugino masses was also analyzed in \cite{nath} 
with interesting results.

The above analyses  
were performed assuming universality (and
non-universality)
of the soft breaking terms at the unification scale, 
$M_{GUT}\approx 10^{16}$ GeV, as it is usually done in the
MSSM literature.
However, 
inspired by superstrings, where the unification scale may be smaller,
it was analyzed in \cite{Nosotros} the sensitivity of the
neutralino-nucleon cross section to the
value of the initial scale for the running
of the soft breaking terms. 
Working in the supergravity context with universal soft terms,
the result was that
the smaller the scale is, the
larger the cross section becomes. In particular, by taking
$10^{10-12}$ GeV rather than $10^{16}$ GeV
for the initial scale, 
the cross section increases substantially 
$\sigma_{\tilde\chi_1^0-p}\approx 10^{-6}$--$10^{-5}$ pb.

The natural extension of this analysis is to carry it out
with
explicit D-brane constructions. 
As mentioned in Subsection 3.1.2,
the first attempts to study dark matter within these constructions
were carried out in 
\cite{khalil}-\cite{Arnowitt} 
for the unification scale as the initial scale
and in \cite{Bailindm} for an intermediate scale as the initial scale
in the case of dilaton dominance.
Here we will take into account the crucial issue of the D-brane origin
of the $U(1)_Y$ and its consequences on the matter distribution 
and soft terms in these scenarios. Thus 
we will analyze the D-brane scenarios introduced
in Section 2, using their soft terms computed in
Section 3.
The fact that in these scenarios
``intermediate'' initial scales and/or non-universal soft terms
are possible allows us to think that large cross sections,
in the small $\tan\beta$ regime,
could be obtained in principle.
Let us recall that this can be understood from
the variation in the value of $\mu$, i.e. the Higgs
mixing parameter which appears in the superpotential $W=\mu H_1 H_2$.
Both, ``intermediate'' scales and  non-universality,
can produce a decrease in the value of $\mu$.
When this occurs, the Higgsino components, $N_{13}$ and $N_{14}$,
of the lightest neutralino
\bea
\tilde\chi_1^0= N_{11}\tilde B^0+N_{12}\tilde W^0
+N_{13}\tilde H_1^0
+N_{14}\tilde H_2^0
\label{neutralino}
\eea
increase
and therefore the scattering channels through Higgs exchange 
become important. As a consequence the
spin independent cross section also increases.

Before entering in details let us remark that
we will work with 
the usual formulas for 
the elastic scattering
of relic LSPs on protons and neutrons that can be found in the 
literature \cite{review}.
In particular, we will follow the re-evaluation of the 
rates 
carried out in \cite{Ellis}, using their 
central values for the hadronic matrix elements.

Let us discuss now the parameter space of our D-brane scenarios.
As usual in supersymmetric theories, the requirement of correct
electroweak breaking leaves us (modulo the sign of $\mu$)
with the following parameters. The soft breaking terms, scalar and gaugino
masses
and trilinear parameters, and $\tan\beta$.
Although formulas for
the soft terms obtained in Section 3 leave us in principle with
five parameters, $m_{3/2}$, $\theta$ and  $\Theta_i$ with $i=1,2,3$,
due to relation $\sum_i |\Theta_i|^2=1$ only four of them are
independent.
In our analysis we vary the parameters $\theta$ and $\Theta_i$
in the whole allowed range, $0\leq \theta < 2\pi$,
$-1\leq\Theta_i\leq 1$. For 
the gravitino mass we take
$m_{3/2}\leq 300$ GeV.
Concerning Yukawa couplings we will fix their values imposing 
the correct fermion mass spectrum at low energies, i.e.
we are assuming that Yukawa structures of D-brane scenarios
give rise to those values.

We will analyze first the scenario of Subsection 2.1.1 with three different
sets of D$p$-branes, where the standard model gauge groups live. 
Since for the third and fourth assignments of
quantum numbers in Table 1 $M_I\lsim 3\times 10^{8}$ GeV, and therefore
$m_{3/2}$ is too low to be phenomenologically interesting,
we will consider only the soft masses associated to the first and second
assignments, i.e. the eight possible combinations given by eqs.
(\ref{gaugino1}-\ref{leptons}), 
(\ref{Higgses1}-\ref{Higgses4}).
The discussion of the corresponding trilinear parameters can be found
below eq.(\ref{Higgses4}).

\begin{figure}[!pht]
\begin{center}
\epsfig{file= 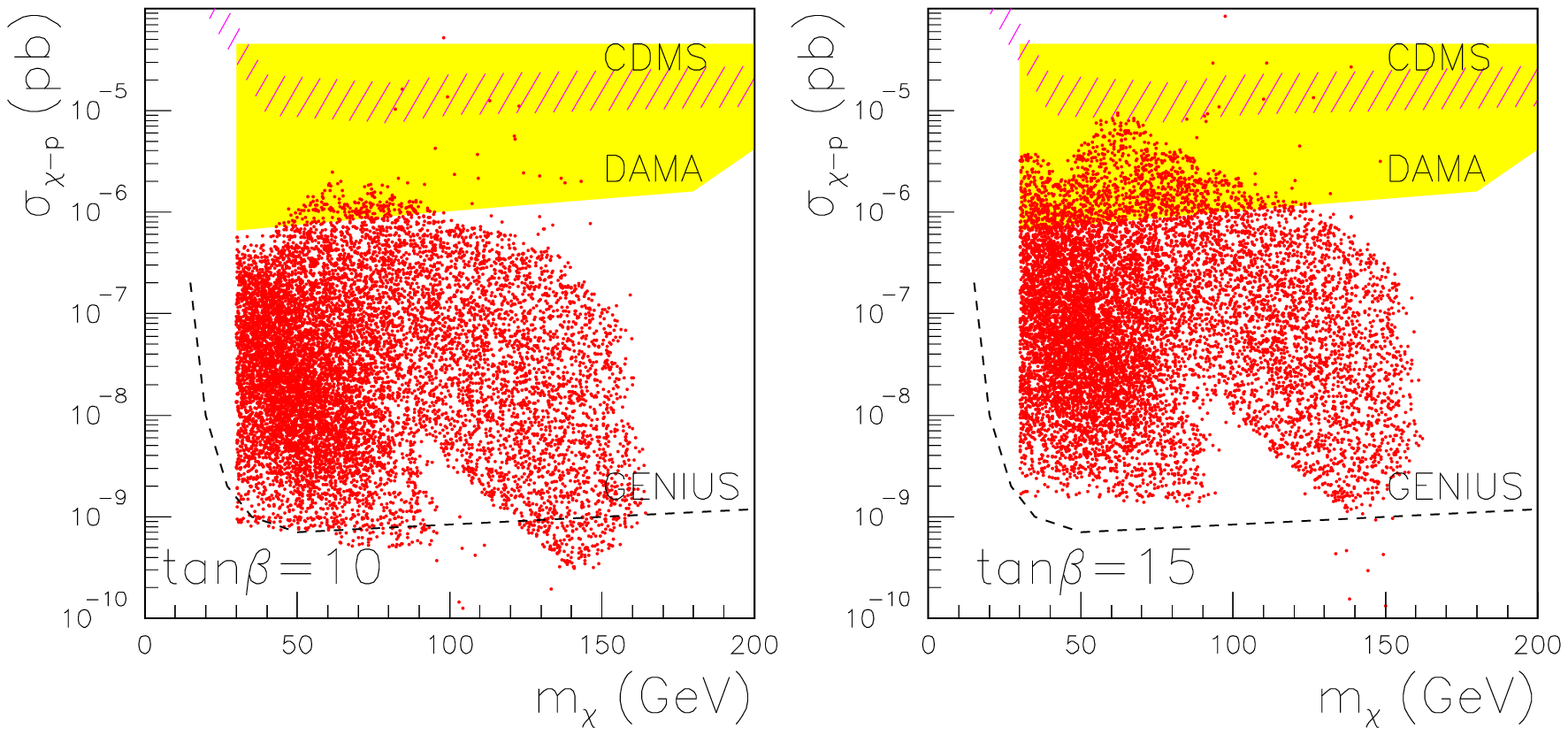, width=16cm
}
\end{center}
\vspace{-1cm} 
\caption{
Scatter plot of the neutralino-proton cross section
as a function of the neutralino mass for the scenario with
three different sets of D$p$-branes. 
The string scale is $M_I=10^{12}$ GeV.
DAMA and CDMS current limits 
and projected GENIUS limits are shown.} 
%
%
\vspace{-1cm}
\begin{center}
\epsfig{file= 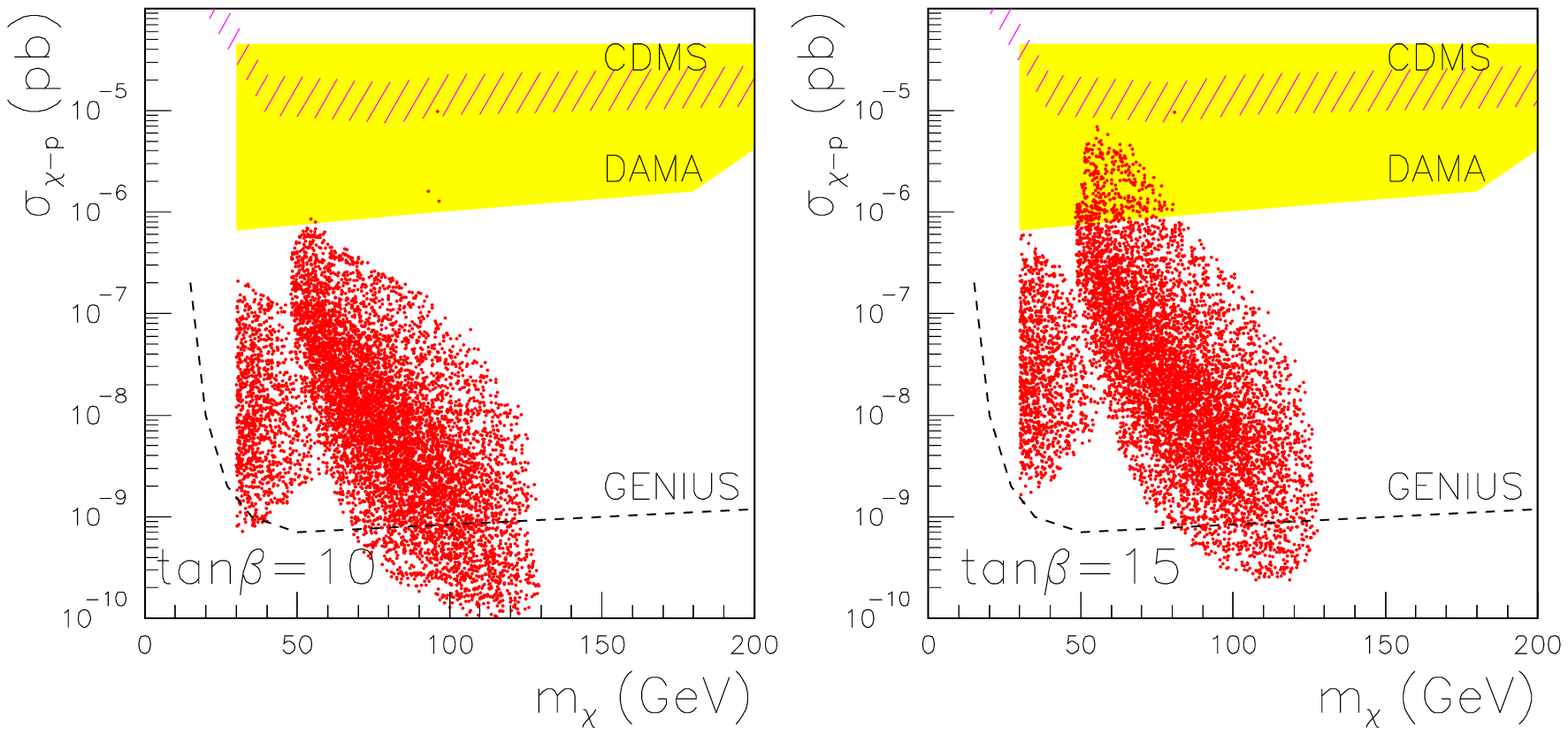, width=16cm
}
\end{center} 
\vspace{-1cm}
\caption{The same as in Fig.~4 but for the string scale 
$M_I=5\times 10^{15}$ GeV.} 
\end{figure}
\begin{figure}[!th]
\begin{center}
\epsfig{file= 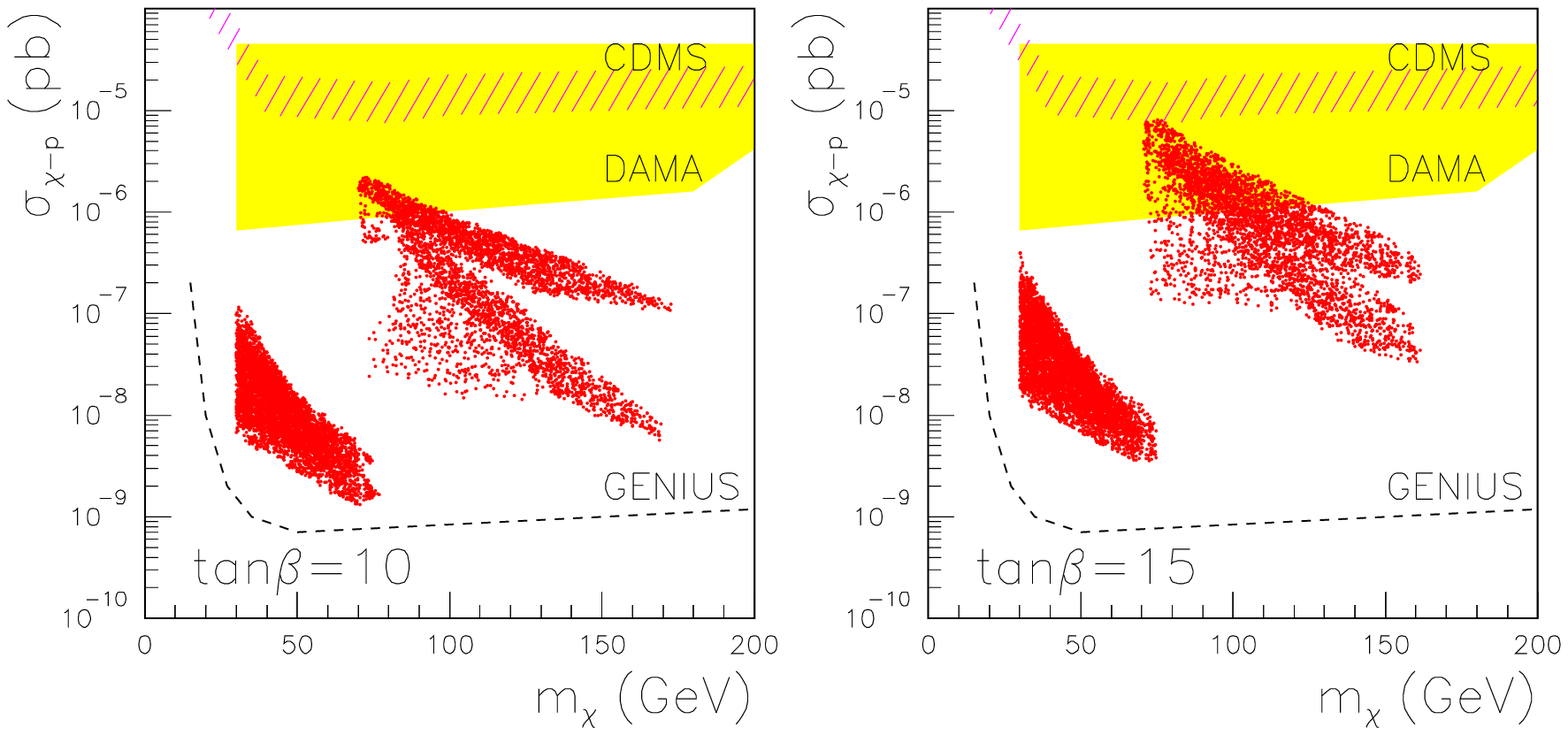, width=16cm
}
\end{center} 
\vspace{-1cm}
\caption{The same as in Fig.~4 but for the scenario
with D$p_1$ = D$p_3$. The string scale is 
$M_I=8\times 10^{9}$ GeV.} 
\end{figure}

In particular, the cross sections associated to combination 
(\ref{gaugino1}), (\ref{scalars1}) and 
(\ref{Higgses1}) are shown in Fig.~4
for $M_I=10^{12}$ GeV (i.e. the example studied in Fig.~2). 
The other possible combinations
give rise to similar results.
Fig~4 displays a scatter plot
of 
$\sigma_{\tilde\chi_1^0-p}$ as a function of the 
LSP mass $m_{\tilde\chi_1^0}$
for a scanning of the parameter space 
discussed above. 
Two different values of $\tan\beta$, 10 and 15, are shown.
Although this and the other figures below have been obtained
using negative values of $\mu$, for positive values
the corresponding figures are equal. Notice that  
the spectrum of supersymmetric particles is invariant under the
transformation $\mu, A, M\rightarrow -\mu, -A, -M$. Since
the shift $\theta\rightarrow \theta + \pi$
implies for the soft terms $M\rightarrow -M$, $A\rightarrow -A$
and $m\rightarrow m$, a figure with positive $\mu$ will be equal  
to a figure with negative $\mu$ shifting $\theta\rightarrow \theta + \pi$.
We have included in the figures LEP and Tevatron bounds
on supersymmetric masses. They forbid e.g. values of
$m_{3/2}$ smaller than 170 GeV.
Although bounds coming from CLEO $b\rightarrow s\gamma$ branching
ratio measurements are not included in the figures, we have checked
explicitly that their qualitative patterns are not modified
when such a bounds are considered.
It is worth noticing that 
for $\tan\beta =10$ there are regions of the parameter
space consistent with DAMA limits. 
In fact, we have checked that $\tan\beta > 5$ is enough to
obtain compatibility with DAMA. 
Since the larger $\tan\beta$ is, the
larger the cross section becomes, for $\tan\beta =15$ these regions
increase.

As discussed below eq.(\ref{MI19}), larger values of the string scale
may be obtained with $\alpha_1(M_I)>0.1$. In particular we show the example
where $M_I=5\times 10^{15}$ GeV, corresponding to 
$\alpha_1(M_I)\approx 1$, in Fig.~5.
Since the larger the scale is, the smaller the cross section becomes,
now the cross sections decrease with respect to the previous case. 
In particular, 
$\tan\beta > 10$ is necessary in order to have compatibility with
DAMA. On the other hand, as discussed above,
in the MSSM with
universal soft terms at the unification scale (which is close to the
above $M_I$),
$\tan\beta\gsim 20$ was needed to obtain compatibility. 
Clearly the non-universality of the
soft terms in this string scenario plays a crucial role
increasing the cross sections.

Let us finally recall that both figures are obtained taking
$m_{3/2}\leq 300$ GeV, which corresponds to squark masses 
$m_{\tilde q}\lsim 500$ GeV  at low energies. We have checked that
larger values of $m_{3/2}$ produce cross sections 
below DAMA limits. In particular, 
the right hand side and bottom of the figures will also be filled
with points.
Cross sections below
projected GENIUS limits will appear in both figures.
On the other hand, it is worth mentioning that the few isolated 
points in the plots with, in general, very large values of the cross section
correspond to values of the lightest stop mass extremely close to the
mass of the LSP, 
in particular $(m_{\tilde t}-m_{\tilde\chi_1^0})/m_{\tilde t}<0.01$.

\begin{figure}[ht]
\begin{center}
\epsfig{file= 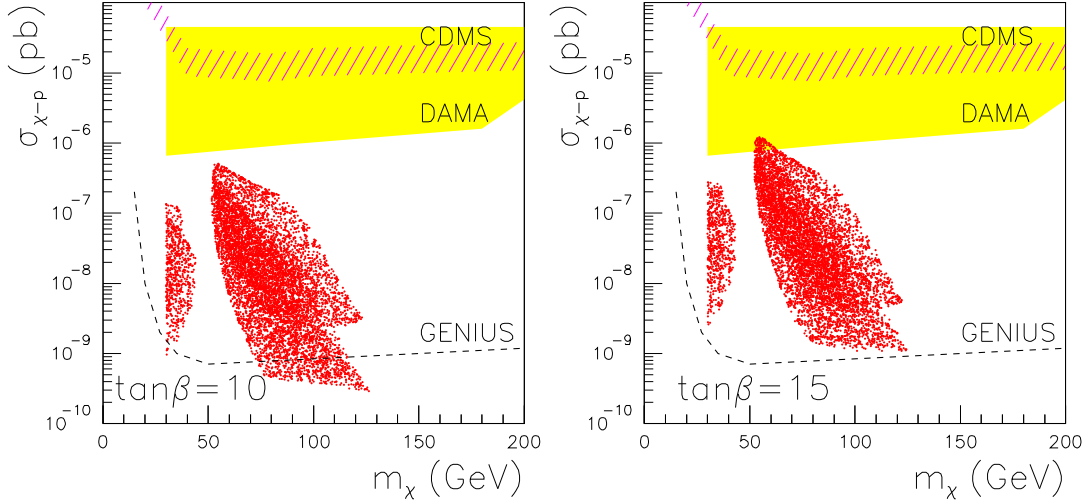, width=16cm
}
\end{center} 
\vspace{-1cm}
\caption{The same as in Fig.~4 but for the scenario
without D$p_1$-brane. The string scale is 
$M_I=10^{16}$ GeV.} 
\end{figure}

Although the scenario of Subsection 2.1.2 where
D$p_1$ = D$p_3$ has soft terms different from the previous scenario 
(see Subsection 3.1.2),
the qualitative results concerning neutralino-proton cross sections
will be similar. This is shown in Fig.~6 
for the example
discussed below eq.(\ref{extra}) where 
$M_I=8\times 10^{9}$ GeV. We use the soft terms given by
(\ref{gauginos1}), (\ref{scalarss1}) and (\ref{Higgsess1})
with the constraint $\Theta_2=\Theta_3$ in order to avoid FCNC,
as discussed in Subsection 3.1.2. Thus, apart from $\tan\beta$, 
only three independent 
parameters are left: $m_ {3/2}$, $\theta$ and one of the $\Theta_i$'s. 
Other combinations of soft terms do not modify our conclusions.
Note that there are regions of the parameter space
consistent with DAMA limits, as in Fig.~4.
In this scenario $\tan\beta > 5$ is also enough to obtain
such a consistency.

The scenario without D$p_1$-brane studied in Subsection 2.1.3
is shown in Fig.~7. We take the string scale
$M_I=10^{16}$ GeV and the soft terms given in Subsection 3.1.3,
with the constraint $\Theta_2=\Theta_3$ to avoid FCNC as discussed there.
Since the string 
scale is large the results are qualitatively similar to the ones
in Fig.~5.

\begin{figure}[ht]
\begin{center}
\epsfig{file= 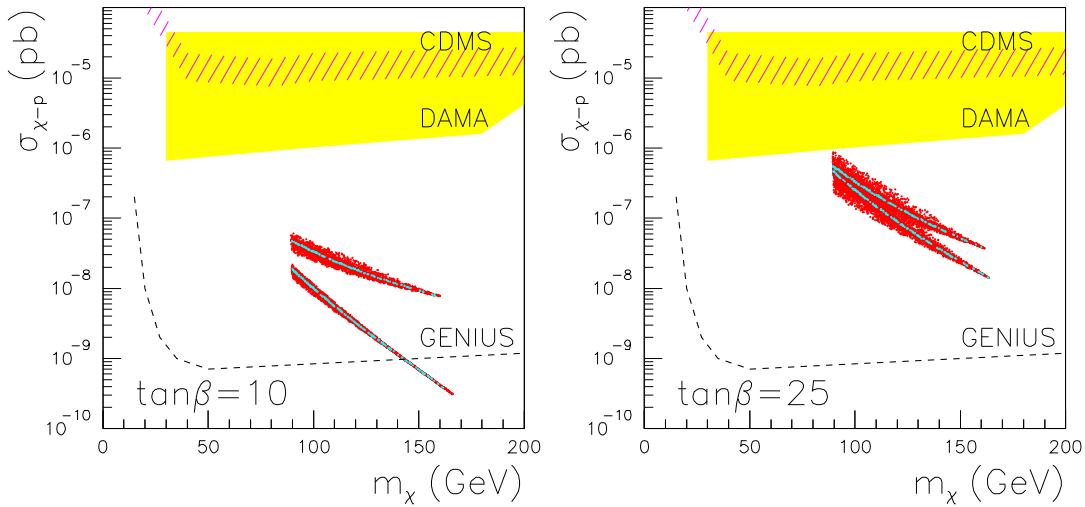, width=16cm}
\end{center} 
\vspace{-1cm}
\caption{The same as in Fig.~4 but for the scenario 
with all gauge groups embedded within the same set of D3-branes,
in such a way that gauge couplings unify at 
$M_I=10^{10}$ GeV.} 
\end{figure}

Let us finally analyze the scenario of Subsection 2.2 
where all gauge groups are embedded within the same set of D3-branes.
In this scenario the soft terms are given by (\ref{gaugino3}),
(\ref{scalars3}), (\ref{masss}) and
(\ref{messs}). Since we will take
$\Theta_1=\Theta_2$ in order to avoid FCNC, there will be in our
analysis
only three independent parameters: $m_{3/2}$, $\theta$ and one of the
$\Theta_i$'s, say $\Theta_3$.
The cross sections are then shown in Fig.~8
for $M_I=10^{10}$ GeV, i.e. the example studied in Fig.~3.
We consider two cases with $\tan\beta = 10$ and $\tan\beta = 25$.
Now $\tan\beta > 25$ is necessary to obtain regions
consistent with DAMA limits.
This is to be compared with the previous scenarios with intermediate string
scales where regions consistent with DAMA were obtained
for $\tan\beta > 5$. 
As discussed in the context of the MSSM in \cite{Nosotros} 
this is due to the different values of $\alpha$'s at the string scale  
in both types of scenarios. Unlike the previous ones 
here gauge couplings are unified.

Let us recall that this scenario is the only one where universality
can be obtained. This is the case for the dilaton limit 
($\sin^2\theta=1$) and the overall modulus limit
($\Theta_{1,2,3}=1/\sqrt 3$). The two central curves in the
figures correspond precisely to this situation. Notice that
the deviation from universality may increase or decrease 
the cross sections, as shown in the figures, depending 
on the values of the parameters $\theta$ and $\Theta_3$ chosen.

Before concluding let us discuss very briefly the effect of relic neutralino 
density bounds on cross sections.
The most robust evidence for the existence of dark matter comes from 
relatively small scales.
Lower limits inferred
from the flat rotation curves of spiral galaxies 
\cite{review,salucci} are 
$\Omega_{halo}\gsim 10\ \Omega_{vis}$ or 
$\Omega_{halo}\ h^2\gsim 0.01-0.05$, where
$h$ is the reduced Hubble constant. 
On the opposite side, observations at large scales,
$(6-20)\ h^{-1} $ Mpc, have provided 
estimates of $\Omega_{CDM} h^2\approx 0.1-0.6$ 
\cite{freedman}, but values as low as 
$\Omega_{CDM} h^2\approx 0.02$ have also been quoted \cite{kaiser}.
Taking up-to-date limits on $h$, the 
baryon density from nucleosynthesis and overall matter-balance  
analysis one is able to obtain a favoured range,
$0.01\lsim \Omega_{CDM} h^2 \lsim 0.3$ (at $\sim 2\sigma$ CL)
 \cite{sadoulet,primack00}.
Note that  conservative 
lower limits in the small and large scales are 
of the same order of magnitude.

In this work the expected neutralino cosmological relic density
has been computed according to well 
known techniques (see \cite{review}). 
Although bounds coming from them
are not included in the above figures we have checked explicitly
that their qualitative patterns are not modified for most of the
regions in figs.~4-7 when such a bounds are considered. 
However the analysis of 
regions of the parameter space consistent with DAMA limits
is more delicate.
From the general behaviour  
$\Omega_{\chi} h^2\propto 1/ \langle \sigma_{ann}\rangle$,
where $\sigma_{ann}$ is the cross section for annihilation of 
neutralinos,
it is expected that such high neutralino-proton cross 
sections as those presented above will then correspond to 
relatively low relic neutralino densities.
We have seen that this is in fact the case. 
On these grounds\footnote{Of course there is always the 
possibility that not all the dark matter in our Galaxy are
neutralinos. This would modify the analysis since e.g. 
$\Omega_{\chi}<\Omega_{CDM}$.}, 
most of those points are
at the border of the range of validity or below.
On the other hand, it is worth remarking that we are clearly below
of the range of validity for the whole regions in the scenario corresponding
to Fig.~8.

\section{Conclusions}

In this paper we have analyzed 
different phenomenological aspects of D-brane scenarios.
First, 
assuming that the 
$SU(3)_c$, $SU(2)_L$ and $U(1)_Y$ groups
of the standard model come from different sets of D$p$-branes,
intermediate values for the string scale $M_I\approx 10^{10-12}$ GeV
are obtained in a natural way. The reason is the following.
Due to the D-brane origin of the $U(1)_Y$ gauge group, the hypercharge
is a linear combination of different $U(1)$ charges. Thus, in order
to reproduce the
low-energy data, i.e. the values of the gauge
couplings deduced from CERN $e^{+}e^{-}$ collider LEP experiments,
intermediate values for $M_I$ are necessary. 
On the other hand, there is also the possibility that
the gauge groups of the standard model come from the same set of
D$p$-branes.
In fact explicit models with this property can be found in the literature.
Again the $U(1)_Y$ gauge group has a D-brane origin and therefore the
normalization factor of the hypercharge is not as the usual one in
GUTs. The presence of additional doublets or triplets allows
to obtain intermediate values for the string scale.

Second, taking into account the matter assignment to the different
D$p$-branes of the above scenarios, we have derived Yukawa couplings and 
soft supersymmetry-breaking terms.
The analysis of the soft terms has been carried out under the
assumption of dilaton/moduli supersymmetry breaking, and they
turn out to be generically non-universal.

Finally, we have computed the neutralino-nucleon cross section
of these D-brane scenarios.
This computation is extremely interesting since
the lightest neutralino 
is a natural candidate
for dark matter in supersymmetric theories.
Using the previously obtained soft terms, and taking into account
that the string scale $M_I$ is the initial scale for their running,
we have found
regions in the parameter space of the D-brane scenarios with
cross sections in the range 
of $10^{-6}$--$10^{-5}$ pb. For instance, this can be obtained for 
$\tan\beta > 5$.
The above mentioned range is precisely the one where 
current dark matter detectors, as e.g. 
DAMA and CDMS,
are sensitive.

\bigskip

\noindent {\bf Acknowledgments}

\noindent 
D.G. Cerde\~no acknowledges the financial support
of the Comunidad de Madrid through a FPI grant.
The work of 
S. Khalil  
was supported by PPARC. 
The work of C. Mu\~noz was supported 
in part by the Ministerio de Ciencia y Tecnolog\'{\i}a, 
and
the European Union under contract HPRN-CT-2000-00148. 
The work of E. Torrente-Lujan was supported in part by  
the Ministerio de Ciencia y Tecnolog\'{\i}a.

\appendix

\makeatletter
\@addtoreset{equation}{section}
\makeatother
\renewcommand{\theequation}{\thesection.\arabic{equation}}

\section{Appendix}

We summarize in this Appendix the formulas for the soft
terms \cite{Rigolin2} and Yukawa couplings \cite{Yukawas}
in D-brane constructions, using 
one set of 9-branes and
three sets of 5-branes, $5_i$.
Assuming vanishing cosmological constant and neglecting phases,
the gaugino masses are given by 
\bea
M_9 & = & \sqrt{3}  m_{3/2} \sin \theta  \ , \nn \\
M_{5_i} & = & \sqrt{3}  m_{3/2}  \Theta_i \cos \theta   \ ,
\label{gaugino111}
\eea
where $M_9$ ($M_{5_i}$) are the masses of gauginos coming from open
strings starting and ending on 9 ($5_i$)-branes.
The scalar masses are given by
\bea
m^2_{\cni} & = & m^2_{\ccjk} = 
m_{3/2}^2\left(1 - 3 \Theta_{i}^2\  \cos^2 \theta \right)
\ , \nn \\
m^2_{\ccii} & = &  m_{3/2}^2\left(1   - 3 \sin^2 \theta \right)
\ , \nn \\
m^2_{\cnci} & = &  m_{3/2}^2\left[1  - \frac{3}{2}   
\left(1 - \Theta_{i}^2 \right)
\cos^2 \theta \right]
\ , \nn \\
m^2_{\ccicj} & = & m_{3/2}^2 \left[1  - \frac{3}{2}  
\left(\sin^2\theta + \Theta_{k}^2 \cos^2\theta  \right)\right] 
\ , 
\label{scalar111}
\eea
where $C^9_i$
denote matter fields coming from open strings starting
and ending on 9-branes (the index $i$ which
labels the three complex compact dimensions is very
useful in order to generate three families as we discuss in the text),
$C^{5_i}_i$ and 
$C^{5_i}_j$ with $i\not= j$ 
are analogous to the previous ones but with 9-branes
replaced by $5_i$-branes,
$C^{95_i}$ 
denote matter fields coming from open strings starting (ending)
on a 9-brane and ending (starting) on a $5_i$-brane,
$C^{5_i5_j}$ with $i\not= j$ 
come from open strings
starting
in one type of $5_i$ brane and ending on a different type 
of $5_j$-brane. 
Finally the results for the trilinear parameters are
\bea
A_{C^9_1 C^9_2 C^9_3} & = & A_{ C_i^{9} C^{95_i} C^{95_i}} =
   -\sqrt{3}  m_{3/2}\sin\theta  
\ , \nn \\
A_{C_1^{5_i}C_2^{5_i}C_3^{5_i}} & = & A_{C_i^{5_i} C^{95_i}C^{95_i}} =
   A_{\ccij \ccick \ccick} = -\sqrt{3}m_{3/2} \Theta _i \cos\theta 
\ , \nn \\
A_{\ccacb \cccca \ccbcc} & = & \frac{\sqrt 3}{2}m_{3/2}
   \left[\sin\theta  - \left(\sum_i\Theta _i \right) \cos\theta \right]
\ , \nn \\
A_{\ccicj \cnci \cncj} & = &  \frac{\sqrt 3}{2}m_{3/2}
   \left[\cos\theta \left(\Theta_k - \Theta_i  
  - \Theta _j \right) - \sin\theta \right] \ ,
\label{trilin1}
\eea
with $i,j,k = 1,2,3$ and $i \neq j \neq k \neq i$ in the above equations.
The angle $\theta$ and the $\Theta_i$ with $\sum_{i} |\Theta_i|^2=1$,
just parameterize the direction of the goldstino in the $S$, $T_i$ field
space.

On the other hand,
the renormalizable Yukawa couplings which are allowed are given by 
\beqa
\cal L_Y & = & g_9 \left(\cna \cnb \cnc \ + \ \ccacb \cccca \ccbcc \ + \ 
\sum_{i=1}^3 \cni \cnci \cnci \right) \ + \  \sum_{i,j,k=1}^3 g_{5_i} 
\left(\ccia \ccib \ccic \right. \nn \\
 & & \ + \ \left. \ccii \cnci \cnci \ + \ d_{ijk} \ccij \ccick \ccick 
\ + \ \frac{1}{2} d_{ijk}  \ccjck \cncj \cnck \right) \ ,
\label{superpottt}
\eeqa
with $d_{ijk}=1$ if $i\neq j\neq k\neq i $, otherwise $d_{ijk}=0$.


\newpage

\end{document}